# Transfer learning discovery of molecular modulators for perovskite solar cells

Authors: Haoming Yan[1,2,7], Xinyu Chen[3,7], Yanran Wang[1,7], Zhengchao Luo[2,3], Weizheng Huang[1], Hongshuai Wang[2], Peng Chen[1], Yuzhi Zhang[2], Weijie Sun[2], Jinzhuo Wang[3]*, Qihuang Gong[1,4,5], Rui Zhu[1,4,5]*, Lichen Zhao[1,6]*

[1]State Key Laboratory of Artificial Microstructure and Mesoscopic Physics, School of Physics, Frontiers Science Center for Nano-optoelectronics & Collaborative Innovation Center of Quantum Matter, Peking University, 100871 Beijing, China

[2]DP Technology, Beijing 100089, China

[3]Department of Big Data and Biomedical AI, College of Future Technology, Peking University, Beijing 100871, China

[4]Collaborative Innovation Center of Extreme Optics, Shanxi University, Taiyuan, Shanxi 030006, China

[5]Key Laboratory for Advanced Optoelectronic Integrated Chips of Jiangsu Province, Peking University Yangtze Delta Institute of Optoelectronics, Nantong, Jiangsu 226010, China

[6]Institute for Advanced Materials and Technology, University of Science and Technology Beijing, Beijing 100083, China

[7]These authors contributed equally to this work

*e-mail: wangjinzhuo@pku.edu.cn (J.W.), iamzhurui@pku.edu.cn (R.Z.), lczhao@pku.edu.cn (L.Z.),

**Abstract:**

The discovery of effective molecular modulators is essential for advancing perovskite solar cells (PSCs), but the research process is hindered by the vastness of chemical space and the time-consuming and expensive trial-and-error experimental screening[1,2]. Concurrently, machine learning (ML) offers significant potential for accelerating materials discovery. However, applying ML to PSCs remains a major challenge due to data scarcity and limitations of traditional quantitative structure-property relationship (QSPR) models[3,4]. Here, we apply a chemical informed transfer learning framework based on pre-trained deep neural networks, which achieves high accuracy in predicting the molecular modulator's effect on the power conversion efficiency (PCE) of PSCs. This framework is established through systematical benchmarking of diverse molecular representations, enabling low-cost and high-throughput virtual screening over 79,043 commercially available molecules. Furthermore, we leverage interpretability techniques to visualize the learned chemical representation and experimentally characterize the resulting modulator-perovskite interactions. The top molecular modulators identified by the framework are subsequently validated experimentally, delivering a remarkably improved champion PCE of 26.91% in PSCs.

## Main

PSCs have emerged as a highly promising photovoltaic technology due to their rapidly advancing PCEs[5]. However, defects within the perovskite film bulk and at interfaces, such as undercoordinated lead ions and halide vacancies, act as non-radiative recombination centers, thereby reducing overall efficiency[6]. Molecular modulators play a crucial role in mitigating these issues by regulating crystallization, passivating defects and suppressing recombination[7]. The chemical space of potential modulators is vast, encompassing a wide variety of functional groups and molecular structures, which makes the identification of optimal candidates highly challenging. Conventional experimental trial-and-error approaches are time-consuming, resource-intensive, and inherently uncertain, rendering exhaustive screening of large molecular libraries impractical. Although density functional theory (DFT) calculations can provide valuable insights into molecule-perovskite interactions, they remain computationally prohibitive for high-throughput exploration of complex systems[8]. Therefore, there is an urgent need for efficient and data-driven strategies to navigate this chemical space and accelerate the discovery of effective modulator molecules for high-performance PSCs[1].

ML has revolutionized materials discovery by enabling data-driven QSPR modeling of molecular systems[9–11]. These models establish correlations between molecular descriptors and material performance, thereby facilitating rapid virtual screening of large molecular libraries[12,13]. Despite these advantages, conventional QSPR approaches face several severe limitations. The scarcity of high-quality experimental data often constrains the model complexity and predictive accuracy[13]. Moreover, the reliance on manually engineered descriptors requires domain expertise and may overlook key chemical features, leading to high dimensionality, multicollinearity, or dependence on computationally expensive quantum calculations[4,8,11,14]. Overfitting frequently occurs when small datasets contain numerous descriptors, resulting in poor generalization to previously unseen molecules[15]. In addition, traditional ML models often struggle to capture the intrinsic complexity of three-dimensional (3D) molecular structures, restricting their applicability to simplified two-dimensional (2D) representations and neglecting steric or conformational effects that are crucial for effective defect passivation in PSCs[15]. Collectively, these limitations hinder predictive reliability and reduce the efficiency of ML-guided experimental validation for passivation effect in PSCs.

Transfer learning has emerged as a powerful paradigm in ML, effectively addressing data scarcity by leveraging models pre-trained on large-scale molecular datasets[16,17]. These deep neural network-based models are trained on millions of molecules to learn general-purpose chemical representations from structures such as Simplified Molecular Input Line Entry System (SMILES) strings, molecular graphs or 3D coordinates[18–20]. Unlike traditional QSPR approaches, they automatically extract rich, task-agnostic features that capture intricate chemical patterns without manual descriptor engineering[21,22]. Pre-training on vast and diverse datasets enables the development of generalized chemical representation for multiple tasks[23]. Subsequent fine-tuning on small, task-specific datasets allows the adaption of this foundational knowledge to predict target properties with high accuracy, even under limited data conditions, while mitigating overfitting[23]. Such models have demonstrated remarkable success across a wide range of chemical and materials science domains, including drug discovery[24], metal-organic frameworks (MOFs)[25], nuclear magnetic resonance (NMR) spectroscopy[26], and organic light-emitting diodes (OLEDs)[27], underscoring their robustness and versatility. Nevertheless, pretrained molecular representation models have not yet been explored for PSC-related tasks, leaving significant potential for innovation in this domain.

In this work, we demonstrate that transfer learning offers a data-efficient strategy for predicting PCE improvement (ΔPCE) in the modulation of PSCs, thereby overcoming the limitations of traditional QSPR methods and accelerating molecular optimization. We first developed a transfer learning framework that fine-tunes pre-trained molecular representation models using a curated dataset of the molecular modulators for PSCs labelled with experimentally measured ΔPCE values. The performance of these pre-trained models was then systematically benchmarked against conventional QSPR models, confirming their superiority predictive accuracy. Specifically, we applied transfer learning to molecular QSPR modeling by fine-tuning state-of-the-art pre-trained models, including Uni-Mol[28], ChemBERTa-2[29], and MolCLR[30]. To ensure comprehensive evaluation, we categorized models into three groups: i. transfer learning models, ii. descriptor-based models employing RDKit- and DFT-calculated molecular features, and iii. fingerprint-based models using representations such as the Klekota-Roth fingerprints (KRFP[31]). Performance comparison across 1D, 2D and 3D input dimensions revealed that the Uni-Mol model featuring SE(3)-equivariant 3D molecular representations, achieved the highest accuracy by effectively capturing spatial characteristics relevant to molecular activity. Guided by this finding, we employed the top-performing Uni-Mol model to perform high-throughput virtual screening across 79,043 commercially available molecules. Interpretability analyses further visualized the learned chemical representations, uncovering key molecular substructures correlated with performance enhancement. Experimental validation confirmed the predicted top-performing molecular modulators, decafluorobenzophenone (DFBP) and 3;4;5;6-tetrachlorophthalonitrile (o-TCPN) which achieved high ΔPCE value of 2.04% and 2.11%, respectively. This integrated workflow establishes a robust and transferable framework for data-driven molecular screening, enabling the rational design of stable and high-efficiency PSCs.

**Overview of the transfer learning framework**

Our framework represents a strategic shift from conventional machine learning approaches, which typically rely on training models from scratch using limited datasets and manually engineered features[32]. Instead, it leverages the knowledge embedded in large-scale pre-trained molecular models and fine-tunes them on a perovskite-specific dataset to identify modulator molecules that enhance the PCE of PSCs (**Supplementary Note 1**). As illustrated in **Fig 1a**, the workflow begins with large-scale self-supervised pre-training on 209 million unlabeled molecules, yielding a foundational SE(3)-equivariant model capable of capturing the intrinsic 3D chemistry of molecular structures, yielding a reusable, generalized, task-unspecific backbone. This transfer-learning strategy directly addresses the issue of data scarcity by reusing latent chemical knowledge. Compared with an otherwise identical model trained from random initialization, the pre-trained model delivers substantially higher accuracy and generalization (**Supplementary Table 1**).

As shown in **Fig. 1b**, the generalized model is then specialized for the perovskite-related task through supervised fine-tuning on a rigorously cleaned dataset of 343 curated modulator molecules with experimentally measured ΔPCE ($\Delta PCE_{exp}$) values. This stage adapts the model to predict ΔPCE and recognize molecule-perovskite interaction pattens. **Supplementary Fig. 1** illustrates the data splitting strategy, which employs a random split, alongside the hyper-parameter tuning and testing framework based on cross-validation.

Furthermore, random splitting was repeated with different random seeds to mitigate the stochastic bias inherent in ML (**Supplementary Fig. 2**). To separate device-dependent baselines from molecule-specific modulation effects, we introduced a linear baseline model relating the initial device PCE to the expected ΔPCE and trained the network to predict the residuals (**Supplementary Note 2 and Supplementary Fig. 3**). A robust data-cleaning method was applied to prevent information leakage (**Supplementary Fig. 4**). The resulting fine-tuned model enables scalable, high-throughput virtual screening (**Fig 1c**), efficiently filters approximately 79,000 commercially available molecules and ranking them according to their predicted ΔPCE ($\Delta PCE_{pred}$). Finally, **Fig. 1d** summarizes the experimental validation of top predicted candidates, comparing the PCE distributions of the control and modulated PSCs and confirming strong agreement between predicted and experimental ΔPCE values.

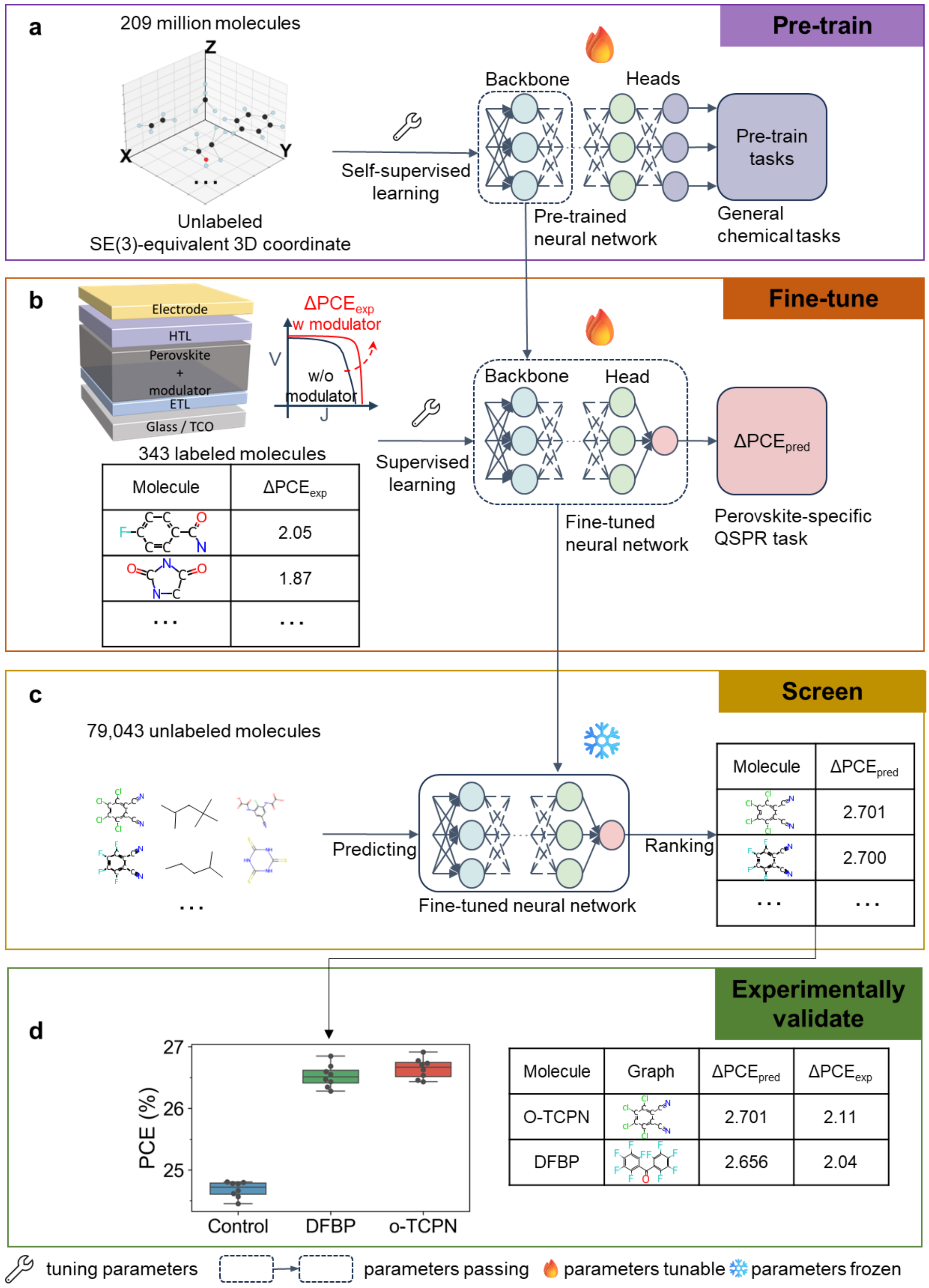

**Fig. 1 | Overview of the transfer learning framework. a**, Self-supervised pre-training on 209 million unlabeled molecules learns general chemical representations. **b**, Supervised

fine-tuning on 343 labeled, perovskite-specific molecules with $\Delta PCE_{exp}$ for QSPR prediction. **c**, High-throughput virtual screening on a dataset of commercially available molecules, and the prediction ranking. **d**, Experimental validation: the statistical box plot of the control, DFBP- and o-TCPN-modulated PSCs shows a significant PCE enhancement for the ML-screened molecules. The right panel: a summary table confirms the consistency between $\Delta PCE_{pred}$ and $\Delta PCE_{exp}$ values.

**Comparative evaluation of conventional QSPR models and transfer learning approaches for PCE prediction**

Our comparative evaluation of different QSPR models begins by categorizing them according to their input representations and learning paradigms. The distinction between conventional ML approaches and modern transfer learning methods is illustrated in **Fig. 2a**. A common type of conventional ML model relies on human-engineered molecular features, typically comprising a dozen or more descriptors These include physicochemical descriptors (e.g., donor number, molecular weight, quantitative estimate of drug-likeness, rotatable bonds, and elemental counts such as C, N, H, F, and O), calculated by RDKit, which serves as 1D or 2D descriptors. The electronic properties such as the highest occupied molecular orbital (HOMO) and the lowest unoccupied molecular orbital (LUMO) gaps calculated via DFT serve as 3D descriptors. Together, these 1D/2D/3D hybrid features contains electronic properties relevant to defect passivation in PSCs. A full list of properties utilized is provided in **Supplementary Table 2**.

Despite their utility, conventional methods face severe limitations. Manual feature engineering can introduce bias and reduce generalization across PSC systems. Features derived from computationally intensive calculations such as DFT-derived properties, significantly hinder high-throughput virtual screening over large molecular libraries. Furthermore, handcrafted descriptors often exhibit substantial multicollinearity (**Supplementary Fig. 5**), which inflates the variance in shallow regressors, whereas learned representations naturally compress correlated signals into task-relevant features. To address this, highly correlated features were removed from the descriptor set. Classical baselines are also sensitive to the number of input features (**Supplementary Fig. 6**). Models tend to underfit with too few descriptors and overfit with too many features, obtaining the best performance with an intermediate feature set. We tested various conventional ML algorithms (**Supplementary Table 3**) and identified Random Forest as the best-performing model for 1D/2D/3D hybrid feature inputs (**Fig. 2b**). The corresponding feature importance and interpretability analysis obtained via SHAP is presented in **Supplementary Fig. 7**.

Another type of conventional ML models employ substructure-based KRFP to represent molecules (**Fig. 2a**). These fingerprints encode 2D graph-like molecular substructures a 4,860-dimensional binary vector, effectively capturing fragment-level features. This representation offers a relatively comprehensive description of key molecular information, such as potential interaction sites with perovskites. While KRFP descriptors are computationally lightweight, they can suffer from information loss in high-dimensional chemical spaces and are limited in capturing holistic molecular properties. Among the algorithms, Gradient Boosting exhibited the best performance when using these 2D fingerprint features (**Fig. 2b**). The corresponding interpretability analysis based on SHAP is presented in **Supplementary Fig. 8**.

Modern deep neural networks enable molecular representation learning through large-scale pretraining on general chemical datasets. The resulting pretrained embeddings can then be effectively transferred to specific photovoltaic tasks for PSCs. ChemBERTa-2 treats molecules as 1D SMILES strings and employs a natural language processing approach based on RoBERTa[33] encoder architecture to extract sequential patterns. By processing 1D SMILES strings through an Attention[34]-based language model, ChemBERTa-2 learns sequential molecular features relevant to chemical properties. MolCLR leverages 2D graph representations of molecules through a pre-trained deep graph neural network (GNN), encoding molecular topology, bond connectivity and atomic environments. This approach effectively captures topological properties but is limited in representing 3D structural features, which can constrain prediction performance of tasks sensitive to spatial conformation. Uni-Mol employs SE(3)-equivariant 3D coordinates to capture full spatial conformations. Its pre-trained neural network backbone enables learning of translationally and rotationally invariant features, providing a robust representation for complex downstream tasks that require detailed 3D structural understanding.

Model performance was then evaluated using metrics including the coefficient of determination ($R^2$) and root mean squared error (RMSE) via 5-fold cross-validation (**Supplementary Fig. 1**), with results averaged over 5 continuous random seeds (**Supplementary Fig. 2**). Additional considerations included inductive bias assessment (**Supplementary Fig. 3**) and duplicate data cleaning (**Supplementary Fig. 4**). The benchmarking results are summarized in **Fig. 2c**,**d** and **Supplementary Table 5**. Among all tested methods, Uni-Mol delivered the highest $R^2$ and the lowest RMSE, demonstrating superior generalization capability of its 3D representation learning. Visualization of predicted versus experimental values (**Fig. 2e-f**) further illustrates the accuracy of Uni-Mol in predicting both ΔPCE and the final PCE. The prediction of ΔPCE holds significant practical value for guiding laboratory experiments, while the final PCE prediction provides an accurate estimate of the ultimate efficiency of PSCs, facilitating straightforward comparisons with similar models reported in the literature.

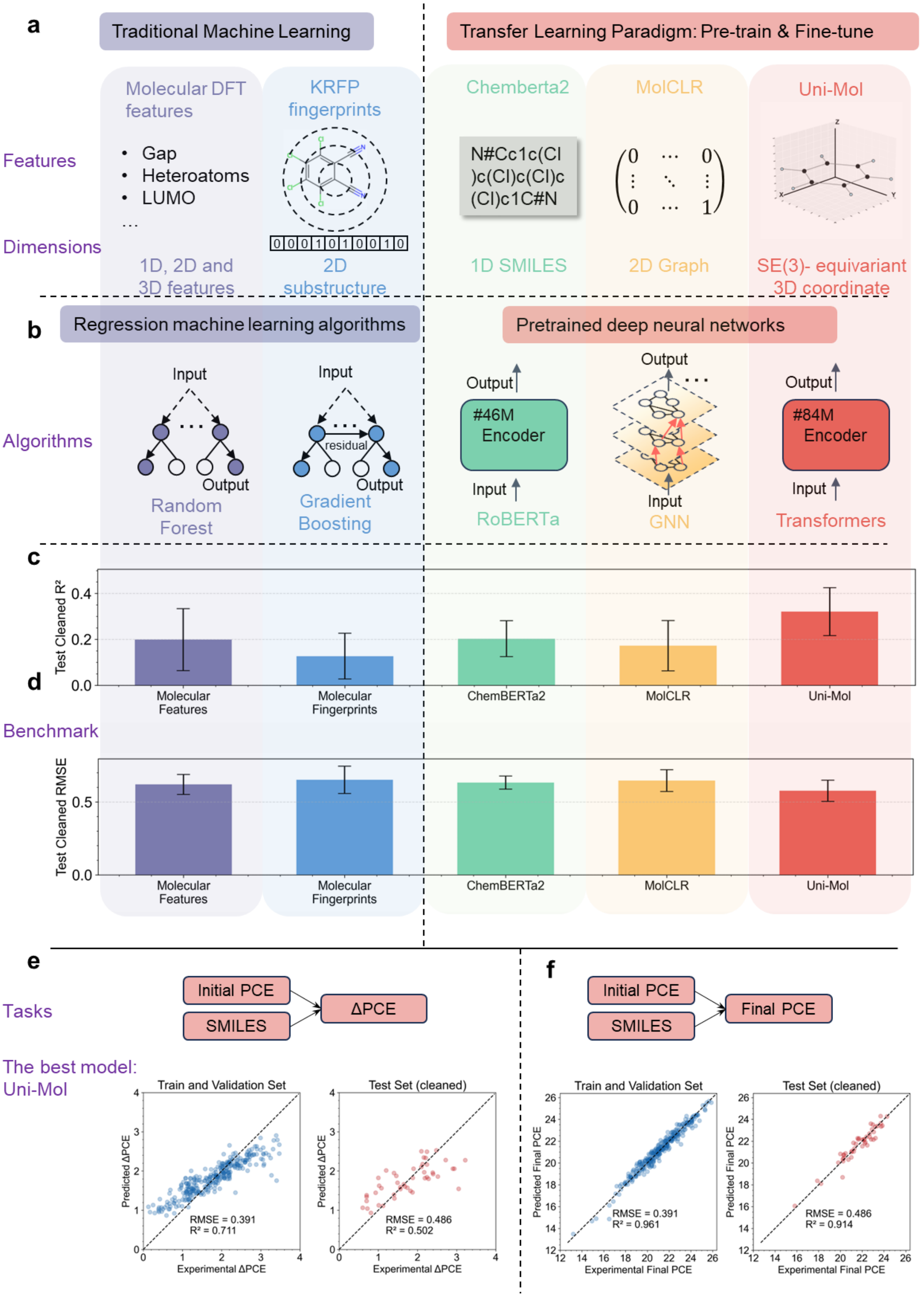

**Fig. 2 | Comparative analysis of ML QSPR molecular representations and model performance. a**, Illustration of input feature representations and molecular information dimensions across conventional and transfer learning paradigms. **b**, Overview of the learning algorithms that achieve the best performance for the corresponding input features. **c-d**, Benchmarking results across all model categories showing $R^2$ scores (**c**, higher is better) and RMSE (**d**, lower is better). Performance metrics were evaluated on the test set and averaged across 5 random splitting seeds. **e-f**, Predictive performance of the best fine-tuned Uni-Mol model for ΔPCE (**e**) and final PCE (**f**) prediction, based on both initial PCE and molecular SMILES input (converted into 3D representation by Uni-Mol). Scatter plots show $\Delta PCE_{pred}$ versus $\Delta PCE_{exp}$ for the training and validation sets (blue points) and the cleaned test set (red points). The dashed diagonal line indicates perfect agreement between predicted and experimental values.

**Large scale virtual screening and interpretation**

Leveraging the superior predictive performance of the fine-tuned Uni-Mol model, we conducted a large-scale virtual screening campaign to identify novel, high-potential molecular modulators from an extensive chemical space. The objective was to move beyond the initial training dataset and discover a diverse set of promising candidates for experimental validation. This approach not only facilitates the identification of top-performing molecules but also enables interpretation of the model's learned principles, providing valuable insights into the underlying QSPR governing perovskite modulation.

Our screening workflow employed a multi-stage filtering protocol to systematically narrow down a massive chemical library into a manageable set of high-priority candidates (**Fig. 3a**). We began with a substantial subset of the PubChem database, selecting 79,043 commercially available molecules from Sigma Aldrich. Molecules with a molecular weight exceeding 400 g $mol^{-1}$ were excluded, resulting in a filtered library of 46,624 candidate molecules. This refined library was then evaluated using our best-performing Uni-Mol model, which rapidly predicted ΔPCE for each molecule, allowing us to rank the entire set and prioritize candidates for further experimental investigation.

The discriminative capability of our fine-tuned Uni-Mol model is clearly demonstrated by comparing the highest-ranked (**Fig. 3b**) and lowest-ranked (**Fig. 3c**) molecules from the virtual screening. The top-10 predicted molecules predominantly feature aromatic cores, such as benzene rings, functionalized with potent electron-withdrawing groups, including cyano (–C≡N) and fluoro (-F) substituents. Both Cyano groups[35] and the π bonds in the halogenated benzene rings[36] have been reported to modulated perovskite materials. These structural motifs are known to promote strong Lewis acid-base interactions with undercoordinated lead ions on the perovskite surface, which is a key mechanism for effective defect modulation. An extended list of the top-100 predicted molecules is provided in **Supplementary Fig. 9**. In contrast, the molecules predicted to be the least effective are generally simple, saturated aliphatic hydrocarbons. These molecules lack the necessary functional groups and electronic properties to interact meaningfully with defect sites in perovskite films, rendering them highly unlikely to act as modulators. This outcome aligns well with human chemical intuition and prior literature, validating the model's ability to capture relevant structure–property relationships.

This distinct chemical differentiation between the top- and bottom-ranked predictions offers robust qualitative validation of the model's autonomously acquired QSPR knowledge. Notably, it demonstrates that, without relying on human-engineered features, the model can effectively capture molecular structures, encompassing both global and local properties, and their correlations with performance, achieving higher predictive accuracy than conventional QSPR approaches.

To move beyond qualitative observations and gain a systematic understanding of the model's feature utilization, we employed advanced interpretability and visualization techniques. Specifically, we analyzed the internal representations encoded within the model's inter-layer weights and employed Uniform Manifold Approximation and Projection (UMAP)[37], a robust dimensionality reduction method, to project the high-dimensional molecular embeddings learned by Uni-Mol into a 2D space amenable to visual inspection and interpretation.

The UMAP visualization of the molecular embeddings provides a compelling demonstration of representation learning capability of Uni-Mol model. As illustrated in **Fig. 3d**, the UMAP visualization of the pre-trained molecular embeddings reveals a dispersed, largely unstructured distribution of data points. The molecules, color-coded according to their predicted effectiveness in PSCs, are extensively intermixed, showing no clear segregation or clustering based on the property of interest. This observation indicates that the initial pre-trained representation, while capturing general molecular features, has not yet optimized the latent space specifically for the regression task on ΔPCE prediction in PSCs.

In contrast, the embeddings from the fine-tuned model (**Fig. 3e**) exhibit a striking and task-relevant reorganization. The data points coalesce into several distinct and well-defined clusters, indicating that the model has successfully adapted its molecular representations to the prediction task for PSCs. This restructured latent space substantially enhances the separability of molecules based on their target properties, with compounds exhibiting similar ΔPCE values (as indicated by the color bar) clustering more closely together in the reduced-dimensional space. The top ten molecules identified through virtual screening were subsequently projected onto the same visualization, shown as red dots in **Fig. 3e**. Notably, all high-priority candidates including those predicted to have the top performance are densely concentrated near these regions, highlighting the model's ability to organize molecules in a task-specific latent space. This clear clustering underscores the effectiveness of fine-tuning in constructing representations that directly facilitate accurate property prediction.

To further validate the distinctions between the molecular datasets corresponding to the left (denoted as Type 1) and right (denoted as Type 2) clusters in **Fig. 3e**, we analyzed their DFT-calculated properties. As illustrated in **Supplementary Fig. 10**, significant differences were observed in several key molecular attributes, whereas other properties (**Supplementary Fig. 11**) exhibited minimal variation. This result confirms that the UMAP clustering of Uni-Mol embeddings effectively captures the intrinsic molecular feature differences, grouping structurally or electronically similar molecules within the same cluster. Hence, the chemical knowledge learned by Uni-Mol is well aligned with the underlying physicochemical properties of the molecules.

The attention heatmap for o-TCPN (**Fig. 3f**) further reveals the atom- and bond-level interpretability of the fine-tuned Uni-Mol model, highlighting the contributions of specific substructures to the predicted molecular activity and interaction with perovskite surface. In particular, the model assigns strong attention weights to the linkages between adjacent carbon atoms on the benzene ring and the cyano group, suggesting these regions as potential interaction sites with undercoordinated $Pb^{2+}$ ions in the perovskite surface regions. The attention heatmap for DFBP, shown in **Supplementary Fig. 12**, exhibits similar trends, reinforcing the model's capacity to capture chemically meaningful interaction patterns. UMAP clustering and attention heatmaps reveal that the model captures both global and atomic-level interaction features, transforming the machine learning black box into a rational design tool.

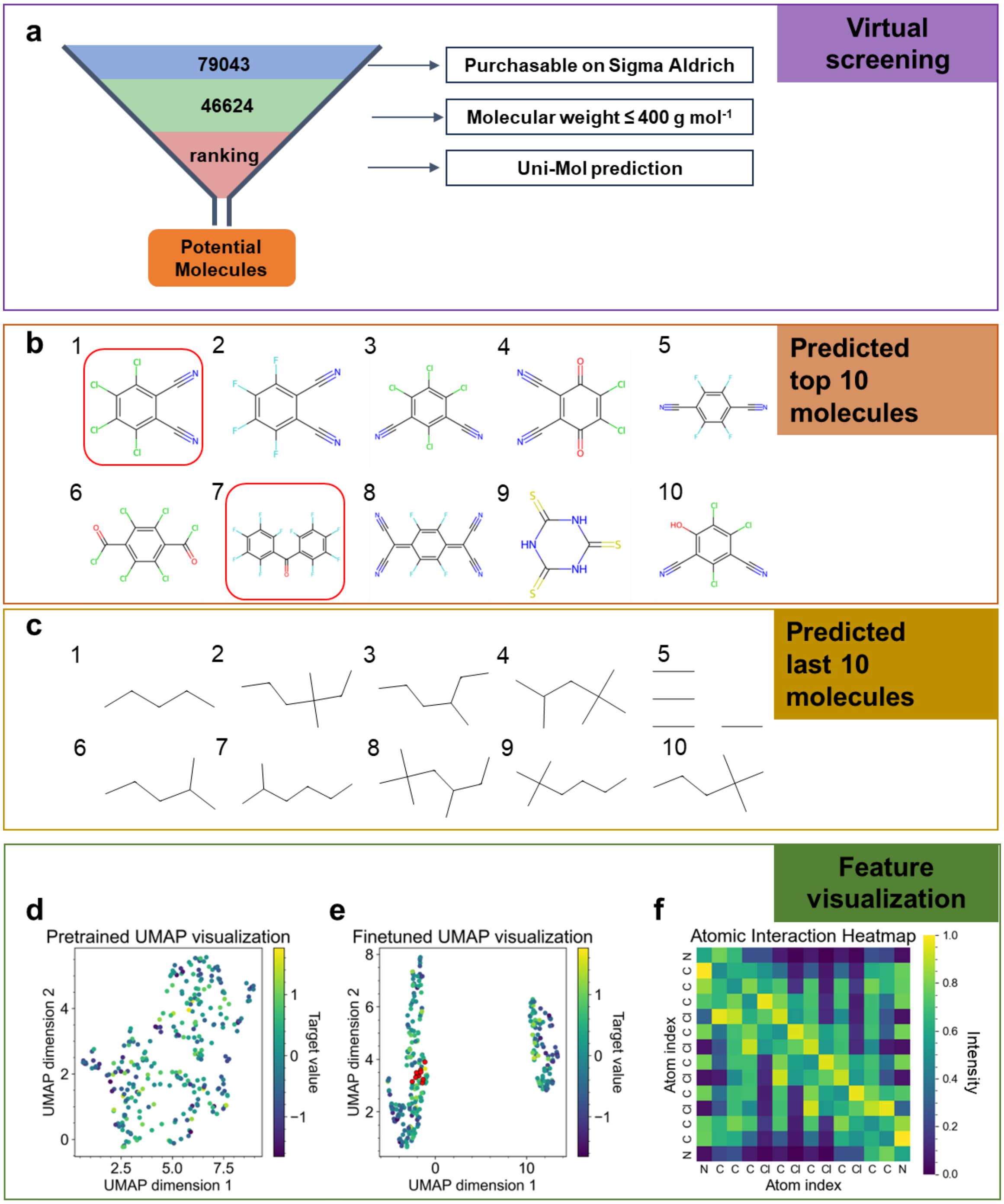

**Fig. 3 | High-throughput virtual screening and model interpretation. a**, Schematic illustration of the three-stage high-throughput virtual screening workflow. **b**, Top 10 modulator molecules predicted by the fine-tuned Uni-Mol model to yield the highest ΔPCE. **c**, Last 10 molecules predicted by the Uni-Mol model, exhibiting negligible or adverse

effects on device performance. **d**-**e**, Visualization of the model's learned representations. UMAP projections of the molecular embeddings before (**d**, pre-training) and after (**e**, fine-tuning) on the PSC-specific dataset, showing the emergence of distinct task-relevant clustering. **f**, Atom-level attention heatmap for o-TCPN molecule screened by the fine-tuned Uni-Mol model. Yellow regions denote stronger atomic correlations and higher contributions to the predicted interaction strength with the perovskite.

To validate the molecules identified through virtual screening, we experimentally investigated their interactions with perovskite in terms of films. Scanning electron microscopy (SEM) was employed to examine morphological variations induced by molecular incorporation. The control film displays evident white $PbI_2$ particulates distributed along the perovskite grain boundaries (**Fig. 4a**). In contrast, the perovskite films incorporated with DFBP (**Fig. 4b**) or o-TCPN (**Fig. 4c**) exhibit markedly improved surface morphology, with $PbI_2$ residues effectively suppressed at the grain boundaries.

The spatially resolved photoluminescence (PL) mapping further elucidates the impact of molecular modulation on the optoelectronic properties of the films. The control film (**Fig. 4d**) exhibits relatively low PL intensity, suggesting abundant non-radiative recombination pathways. Upon the incorporation of DFBP (**Fig. 4e**) or o-TCPN (**Fig. 4f**), a pronounced enhancement in PL intensity is observed, confirming substantial defect passivation and suppression of non-radiative losses. Among them, the DFBP-treated film exhibits the highest PL intensity, while the o-TCPN-treated film also shows a marked improvement in PL intensity relative to the control film.

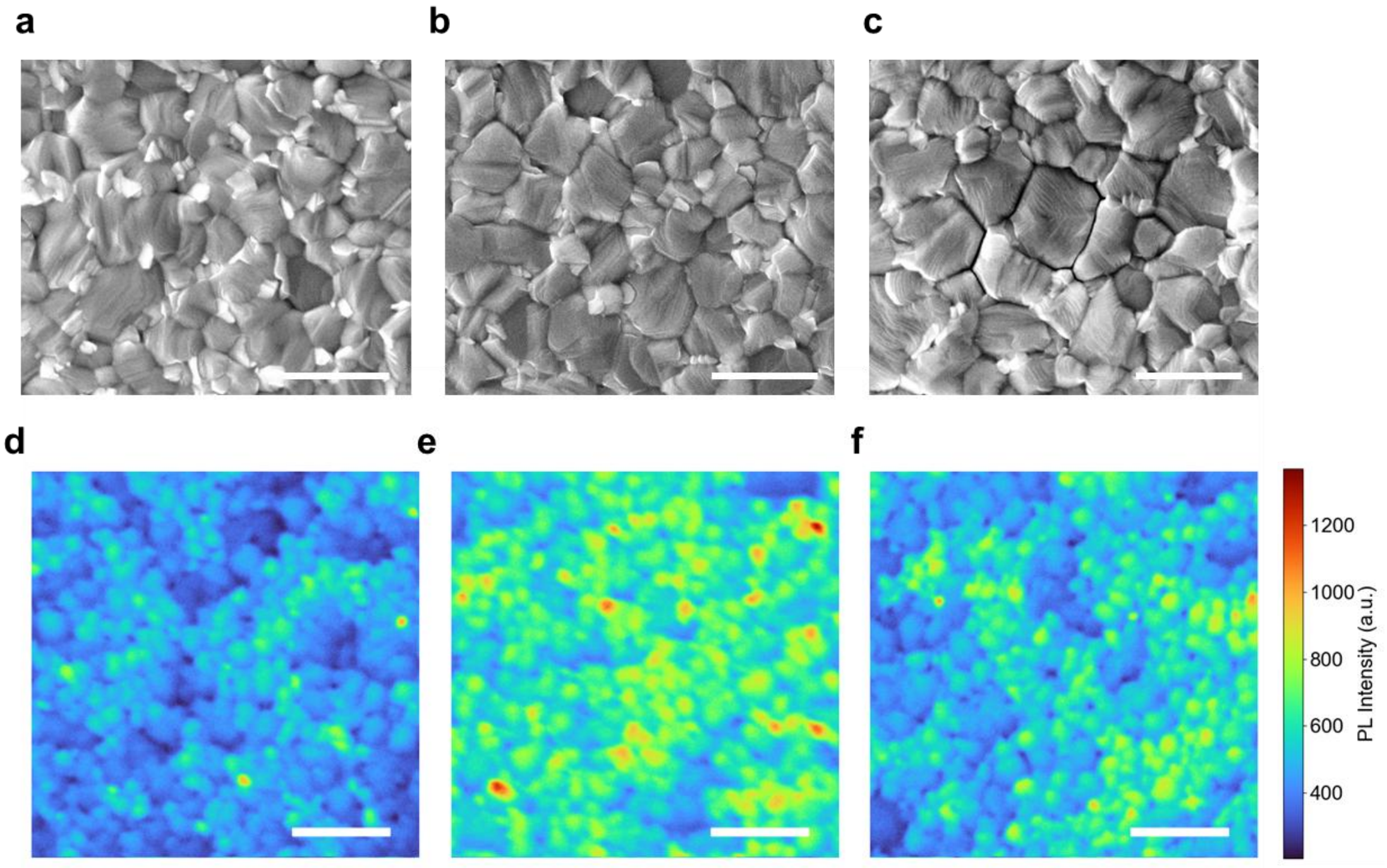

**Fig. 4 | Experimental characterizations of perovskite films. a**-**c**, Top-view SEM images for the control (**a**), DFBP- (**b**) and o-TCPN- (**c**) treated perovskite films. **d**-**f**, Spatial PL intensity mappings for the control (**d**), DFBP- (**e**) and o-TCPN- (**f**) treated films. Scale bars, 2 μm (a-c); 5 μm (d-f).

Current density–voltage (*J*–*V*) analyses (**Fig. 5a**-**c**) confirm that ML-identified o-TCPN delivers the highest photovoltaic performance (a champion PCE of 26.91% for reverse scan) with minimal hysteresis. Detailed photovoltaic statistics are shown in **Supplementary Fig. 13**. Furthermore, the stabilized power output (SPO) measured at maximum power points (MPPs) (**Fig. 5d**) reveal that the molecular modulators markedly enhance the short-term operational stability, maintaining 95%, 98%, and 100% of the initial power output at the end of the measurement period for the control, DFBP-, and o-TCPN-treated devices, respectively.

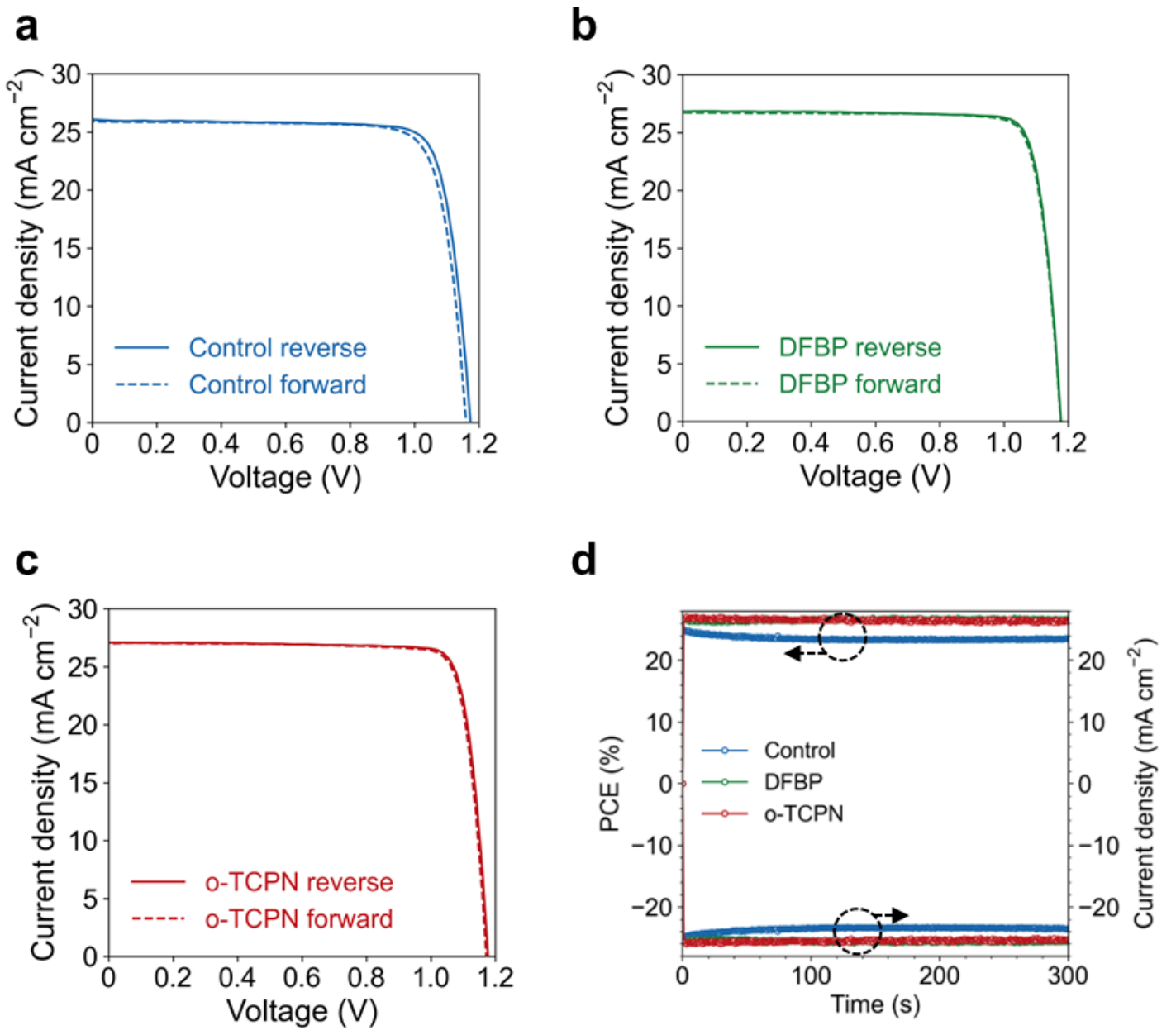

**Fig. 5 | Photovoltaic performance of PSC devices. a-c**, *J–V* curves (forward and reverse scans) of the control (**a**), DFBP- (**b**) and o-TCPN-treated (**c**) PSCs under simulated AM1.5G illumination. **d**, Stabilized power output (SPO) and corresponding operating current density over time at the respective maximum power points (MPPs). The applied voltage biases for the control, DFBP- and o-TCPN-treated PSCs were set to 1.00 V, 1.04 V and 1.04 V, respectively, with a measurement duration of 300 seconds.

## Discussions

This work establishes an end-to-end workflow integrating large-scale pre-training, data-efficient fine-tuning, high-throughput virtual screening, and experimental validation, offering a blueprint for AI-driven discovery of functional materials for next-generation photovoltaics. The key innovation is the first application of a transfer learning-based deep neural network to QSPR modeling of molecule-perovskite interactions. By leveraging the pre-trained Uni-Mol model, our framework transcends the limitations of traditional ML methods reliant on scarce data and hand-crafted descriptors. Systematic benchmarking confirms the superiority of the SE(3)-equivariant Uni-Mol model, underscoring the necessity of incorporating molecular conformations for accurate modulation prediction. A comparative summary of existing ML-based QSPR studies in PSCs is presented in **Supplementary Table 5**. The model also provides interpretable insights which is consistent with chemical intuition.

Our framework is highly scalable and upgradable, allowing the core pre-trained model to upgrade to leverage the future advancements in chemical representation learning society. This technique show great promise in the virtual screening of unlabeled molecules at a large scale with high accuracy, fundamentally accelerating materials discovery. Prioritizing a small, high-potential subset for modulator molecules dramatically reduces the cost and time associated with experimental trial-and-error, thereby accelerates the research cycle in PSCs development and forms a closed loop in materials discovery. The development of such data-driven strategies will significantly shorten the research and development cycle for new and improved PSCs materials, paving the way for a more rational, efficient, and accelerated path toward next-generation solar energy technologies.

**Acknowledgments**

This work was financially supported by the National Natural Science Foundation of China (52325310, U24A6003, 52203208, 52403369, 6220071694), the National Key Research and Development Program of China (2024YFF0507404), the National Key R&D Program of China (2021YFB3800101), Beijing Nova Program (contract no. 20230484480), Young Elite Scientists Sponsorship Program by CAST (YESS20240571),Yunnan Provincial Science and Technology Project at Southwest United Graduate School (202302AO370013), and the R&D Fruit Fund (20210001).

**Author contributions**

All authors analyzed the data, reviewed and commented the paper.

**Conflict of interest**

The authors declare no competing interests.

**Data availability**

Pubchem database available at https://pubchem.ncbi.nlm.nih.gov/

Sigma-aldrich www.sigmaaldrich.cn

Part of the training data was from the perovskite database project[38] https://perovskitedatabase.com/.

**Code availability**

The code, model, and data will be made publicly available at https://github.com/newtontech/Perovskite_Pretrain_Models under an open-source license. An online notebook describing the fine-tuning process will be made available at https://www.bohrium.com/notebooks/34713312597/. An online service app for predicting molecules using our method will be made available at https://www.bohrium.com/apps/perovskite-predict.

**Additional information**

Correspondence and requests for materials should be addressed to the corresponding authors.

## References

1. Yao, Z. *et al.* Machine learning for a sustainable energy future. *Nat. Rev. Mater.* **8**, 202–215 (2022).

2. Chen, P. *et al.* The promise and challenges of inverted perovskite solar cells. *Chem. Rev.* **124**, 10623–10700 (2024).

3. Liu, Y. *et al.* Machine learning for perovskite solar cells and component materials: Key technologies and prospects. *Adv. Funct. Mater.* **33**, 2214271 (2023).

4. Yan, G., Tang, H., Shen, Y., Han, L. & Han, Q. AI-Generated ammonium ligands for high-efficiency and stable 2D/3D heterojunction perovskite solar cells. *Adv. Mater.* 2503154 (2025) doi:10.1002/adma.202503154.

5. Green, M. A. *et al.* Solar Cell Efficiency Tables (Version 65). *Prog. Photovolt. Res. Appl.* **33**, 3–15 (2024).

6. Luo, D., Su, R., Zhang, W., Gong, Q. & Zhu, R. Minimizing non-radiative recombination losses in perovskite solar cells. *Nat. Rev. Mater.* **5**, 44–60 (2019).

7. Gao, F., Zhao, Y., Zhang, X. & You, J. Recent progresses on defect passivation toward efficient perovskite solar cells. *Adv. Energy Mater.* **10**, 1902650 (2020).

8. Xu, J. *et al.* Anion optimization for bifunctional surface passivation in perovskite solar cells. *Nat. Mater.* **22**, 1507–1514 (2023).

9. Chen, Z. *et al.* Machine learning will revolutionize perovskite solar cells. *The Innovation* **5**, 100602 (2024).

10. Zhang, Q. *et al.* Machine-learning-assisted design of buried-interface engineering materials for high-efficiency and stable perovskite solar cells. *ACS Energy Lett.* **9**, 5924–5934 (2024).

11. Zhang, X. *et al.* Machine learning for screening small molecules aspassivation materials for enhanced perovskite solar cells. *Adv. Funct. Mater.* **34**, 2314529 (2024).

12. Wu, J. *et al.* Inverse design workflow discovers hole-transport materials tailored for perovskite solar cells. *Science* **386**, 1256–1264 (2024).

13. Pu, Y. *et al.* Data-driven molecular encoding for efficient screening of organic additives in perovskite solar cells. *Adv. Funct. Mater.* https://doi.org/10.1002/adfm.202506672 (2025) doi:10.1002/adfm.202506672.

14. Huang, Y. & Zhang, L. Descriptor design for perovskite material with compatible molecules via language model and first-principles. *J. Chem. Theory Comput.* **20**, 6790–6800 (2024).

15. Wu, Y. *et al.* Universal machine learning aided synthesis approach of two-dimensional perovskites in a typical laboratory. *Nat. Commun.* **15**, 138 (2024).

16. Zhang, Q. *et al.* Scientific large language models: A survey on biological & chemical domains. Preprint at https://doi.org/10.48550/arXiv.2401.14656 (2024).

17. Xia, J., Zhu, Y., Du, Y. & Li, S. Z. A Systematic survey of chemical pre-trained models. Preprint at https://doi.org/10.48550/ARXIV.2210.16484 (2022).

18. Zheng, Y. *et al.* Large language models for scientific discovery in molecular property prediction. *Nat. Mach. Intell.* https://doi.org/10.1038/s42256-025-00994-z (2025) doi:10.1038/s42256-025-00994-z.

19. Reiser, P. *et al.* Graph neural networks for materials science and chemistry. *Commun. Mater.* **3**, 93 (2022).

20. Liu, S. *et al.* Pre-training molecular graph representation with 3D geometry. Preprint at https://doi.org/10.48550/ARXIV.2110.07728 (2021).

21. Ross, J. *et al.* Large-scale chemical language representations capture molecular structure and properties. *Nat. Mach. Intell.* **4**, 1256–1264 (2022).

22. Qiao, J. *et al.* Molecular pretraining models towards molecular property prediction. *Sci. China Inf. Sci.* **68**, 170104 (2025).

23. Shoghi, N. *et al.* From molecules to materials: Pre-training large generalizable models for atomic property prediction. Preprint at https://doi.org/10.48550/ARXIV.2310.16802 (2023).

24. Tropsha, A., Isayev, O., Varnek, A., Schneider, G. & Cherkasov, A. Integrating QSAR modelling and deep learning in drug discovery: the emergence of deep QSAR. *Nat. Rev. Drug Discov.* **23**, 141–155 (2024).

25. Wang, J. *et al.* A comprehensive transformer-based approach for high-accuracy gas adsorption predictions in metal-organic frameworks. *Nat. Commun.* **15**, 1904 (2024).

26. Xu, F. *et al.* Toward a unified benchmark and framework for deep learning-based prediction of nuclear magnetic resonance chemical shifts. *Nat. Comput. Sci.* **5**, 292–300 (2025).

27. Cheng, Z. *et al.* Automatic screen-out of Ir(III) complex emitters by combined machine learning and computational analysis. *Adv. Opt. Mater.* **11**, 2301093 (2023).

28. Zhou, G. *et al.* Uni-Mol: A universal 3D molecular representation learning framework. in (ICLR, 2023). doi:10.26434/chemrxiv-2022-jjm0j-v4.

29. Ahmad, W., Simon, E., Chithrananda, S., Grand, G. & Ramsundar, B. ChemBERTa-2: Towards chemical foundation models. Preprint at https://doi.org/10.48550/arXiv.2209.01712 (2022).

30. Wang, Y., Wang, J., Cao, Z. & Barati Farimani, A. Molecular contrastive learning of representations via graph neural networks. *Nat. Mach. Intell.* **4**, 279–287 (2022).

31. Klekota, J. & Roth, F. P. Chemical substructures that enrich for biological activity. *Bioinformatics* **24**, 2518–2525 (2008).

32. Chen, M. *et al.* Application of machine learning in perovskite materials and devices: A review. *J. Energy Chem.* **94**, 254–272 (2024).

33. Liu, Y. *et al.* RoBERTa: A robustly optimized BERT pretraining approach. preprint at https://doi.org/10.48550/arXiv.1907.11692 (2019).

34. Vaswani, A. *et al.* Attention is all you need. in *Proceedings of the 31st International Conference on Neural Information Processing Systems* 6000–6010 (Curran Associates Inc., Red Hook, NY, USA, 2017).

35. Nurzhanov, M. *et al.* Dicyanobenzene passivated perovskite solar cells with enhanced efficiency and stability. *J. Mater. Chem. C* **11**, 15881–15891 (2023).

36. Huang, Z. *et al.* Anion–π interactions suppress phase impurities in $FAPbI_3$ solar cells. *Nature* **623**, 531–537 (2023).

37. McInnes, L., Healy, J. & Melville, J. UMAP: Uniform manifold approximation and projection for dimension reduction. Preprint at https://doi.org/10.48550/arXiv.1802.03426 (2020).

38. Jacobsson, T. J. *et al.* An open-access database and analysis tool for perovskite solar cells based on the FAIR data principles. *Nat. Energy* **7**, 107–115 (2021).

## Supplementary Information

### Supplementary Note 1: Machine learning methods

### Pre-train

We directly used pre-trained weights from the Uni-Mol[1] work https://huggingface.co/dptech/Uni-Mol-Models/blob/main/mol_pre_all_h_220816.pt. The model with hydrogen was used.

The pretrained MolCLR pre-trained weights was downloaded from https://github.com/yuyangw/MolCLR.

The pretrained ChemBERTa-2 weight was downloaded from https://huggingface.co/DeepChem/ChemBERTa-77M-MTR.

For randomized pre-training tasks (no pre-training), we used random weights to initialize and compare with the pre-trained neural network. All parameters of the model are randomly sampled from a normal distribution with a mean of 0 and a standard deviation of 0.1.

### The pre-train details of Uni-Mol

Uni-Mol[1] was built upon a Pre-LayerNorm Transformer backbone specifically designed for 3D molecular data. The model maintains two parallel, SE(3)-invariant representations throughout the network: an atom-level representation $\mathbf{h}_i^{(0)} \in \mathbb{R}^d$ initialized from atom type embeddings, and a pair-level representation $\mathbf{p}_{ij}^{(0)} \in \mathbb{R}^d$ initialized from spatial positional encodings. This encoding is derived from the pairwise Euclidean distances $r_{ij} = \left|\left|\boldsymbol{x}_i - \boldsymbol{x}_j\right|\right|$, using a Gaussian radial basis function:

$$\mathbf{p}_{ij}^{(0)} = MLP\left(\left[\exp\left(-\gamma\left(r_{ij} - \mu_k\right)^2\right)\right]_{k=1}^{K}\right)$$

where $\mu_k$ are learnable kernel centers and $\gamma$ is a bandwidth parameter. This explicit 3D encoding is propagated through the network via atom-to-pair and pair-to-atom communication, enabling the model to capture long-range 3D interactions.

The pretraining is performed on a massive dataset of 209 million unlabeled 3D molecular conformations. Two self-supervised tasks are employed. The first is Masked Atom

Prediction (MAP), where 15% of atom types are randomly masked. The model predicts the original atom type $ai$ for each masked atom $i$ , optimized with the cross-entropy loss:

$$L_{atom} = -\sum_{i \in M} logP\left(a_i \middle| h_i^{(L)}\right)$$

where M is the set of masked atoms and $h_i^{(L)}$ is the final atom representation.

The second task is 3D Position Recovery (3D-PR), which directly regresses the true 3D coordinates of a subset of atoms whose positions have been corrupted with uniform noise ($\delta$~U(−1A˚,+1A˚) ). The model's SE(3)-equivariant head predicts a delta-position $\Delta x_i$ . The loss is the mean squared error:

$$L_{coord} = \frac{1}{|\mathrm{N}|} \sum_{i \in N} \left|\left|\Delta x_i - \left(x_i^{true} - x_i^{noisy}\right)\right|\right|^2$$

where N is the set of corrupted atoms.

To ensure training stability, especially in mixed-precision mode, a Representation Normalization Loss is introduced. This loss prevents the intermediate representations from becoming too large or too small, which can cause numerical instability. For any representation $s_i$ (atom or pair), the loss is:

$$L_{norm} = \frac{1}{N} \sum_{i=1}^{N} \max\left(\left||s_i| - \sqrt{d}\right| - \tau, 0\right)$$

where $d$ is the representation dimension, $N$ is the number of representations, and $\tau$=1 is a tolerance factor. This loss is weighted by 0.01 in the total objective. The total pretraining loss is a weighted sum:

$$\mathrm{L}_{total} = L_{atom} + \lambda_{coord} L_{coord} + \lambda_{norm} L_{norm}$$

This comprehensive pretraining framework enables Uni-Mol to learn a universal, transferable 3D molecular representation.

**Fine-tune**

For the Uni-Mol model, we used the unimol_tools python package https://pypi.org/project/unimol-tools/ to fine-tune. Fine-tune process run on A100. We used Optuna[2] to determine the best hyperparameters. The model was trained for 350 epochs with an early stopping patience of 60 epochs, meaning training would halt if the validation loss failed to improve for 60 consecutive epochs. Other key hyperparameters are listed in the table below:

| **Hyperparameter** | **Definition** | **Value/Setting** |
|---|---|---|
| amp | Mixed precision | True |
| anomaly_clean | Anomaly cleaning | True |
| batch_size | Batch size | 1 |
| learning_rate | Learning rate | 8.5e-05 |
| max_norm | Gradient clipping | 12.0 |
| warmup_ratio | Warmup ratio | 0.03 |

For the ChemBERTa-2 model and MolCLR model, All parameters were set to their default values as defined in the project library.

The fine-tuning procedure utilizes the Mean Squared Error (MSE) as the objective function. It is given by:

$$L_{MSE} = \frac{1}{n}\sum_{i=1}^{n}(R(X_i) - f_{\mathrm{QSPR}}(X_i))^2$$

Here, $R(X_i)$ represents the molecule-specific residual term , and $f_{\mathrm{QSPR}}(X_i)$represents the prediction of QSPR model of the i-th molecule.

**Evaluation metrics**

RMSE quantifies the average magnitude of prediction errors in an ML model, providing a measure of how well the model's outputs match actual values. In the context of perovskite solar cells, it can evaluate how accurately an ML algorithm predicts PCE based on input features like perovskite composition or deposition methods. It is calculated as:

$$RMSE = \frac{1}{n}\sum_{i=1}^{n}(y_i - \hat{y}_i)^2$$

Here, RMSE represents the standard deviation of the residuals (i.e., the differences between $\Delta PCE_{pred}$ and $\Delta PCE_{exp}$). Lower RMSE values indicate better predictive accuracy, which is essential for reliable ML-driven screening of perovskite materials.

Before computing $R^2$, we define the baseline reference y-, which is the average of all observed PCE values across the dataset. This serves as a simple benchmark for comparison. It is given by:

$$\bar{y} = \frac{1}{n}\sum_{i=1}^{n} y_i$$

$R^2$ is then calculated as:

$$R^2 = 1 - \frac{\sum_{i=1}^{n}(y_i - \hat{y}_i)^2}{\sum_{i=1}^{n}(y_i - \bar{y})^2}$$

In perovskite solar cell ML models, $R^2$ measures the proportion of variance in the target variable $\Delta PCE_{pred}$ that the model can explain using input features. For a normal predicting ML model, $R^2$ ranges from 0 to 1, where values approaching 1 signify a strong model fit to the data, implying the ML system effectively captures underlying patterns for high-performance cell predictions, values approaching 0 mean total randomness and the model learns nothing. This metric is particularly useful for assessing model reliability in diverse datasets with experimental variability, but it should be paired with domain-specific validations to ensure it drives meaningful materials discovery.

**Benchmarking**

To ensure robust and unbiased performance evaluation, we employed multiple independent random seeds across all experimental configurations. Hyperparameter optimization represents a critical determinant of model performance. We utilized Optuna[2] to systematically optimize hyperparameters through a structured search process. For each random seed, we conducted 20 optimization trials using only the training and validation sets from a 5-fold cross-validation framework. The optimal hyperparameters and corresponding fine-tuned model weights were determined based on validation set performance. Subsequently, these optimized configurations were applied to the test set to obtain final performance metrics.

**Supplementary Note 2: The Inductive Bias-Informed Modeling Framework for Predicting PCE Enhancement**

**Symbol Definitions:**

X: The molecular representation of the additive, typically a SMILES string or other molecular fingerprint onlinelibrary.wiley.com.

$I \in \mathbb{R}$: The initial PCE of the PSC before the additive is introduced.

$P \in \mathbb{R}$: The final PCE of the PSC after the additive is introduced.

ΔPCE: The measured enhancement in efficiency, calculated as ΔPCE = P - I.

Linear ΔPCE(I): The baseline efficiency enhancement, represented as a linear function of I.

R(X): The residual ΔPCE, hypothesized to be dependent on the molecular properties of the additive X.

$\hat{R}(X)$: The predicted residual enhancement from the QSPR model.

**Inductive Bias-Informed Modeling**

Predicting the ΔPCE upon the introduction of a molecular additive to PSC is a primary objective in materials discovery. The conventional approach in QSPR modeling assumes that the ΔPCE is predominantly a function of the additive's molecular structure. However, our analysis of a large dataset aggregated from over 300 publications reveals a strong, systematic trend that challenges this assumption.

As illustrated in the provided scatter plots, we observe a distinct negative linear correlation between I and ΔPCE. We decompose ΔPCE into a linear baseline dependent on I and a molecule-specific residual R(X). This trend, which constitutes a powerful inductive bias, suggests that the performance of the starting device significantly constrains the potential for improvement. Our proposed modeling framework is designed to explicitly incorporate this statistical prior, thereby separating the system-dependent baseline effect from the intrinsic, molecule-specific contribution. This is achieved by modeling a residual value rather than the raw ΔPCE. To formalize this approach, we define a framework based on the following components, hypotheses, and learning objectives.

To ensure unbiased generalization, we evaluate on a deduplicated, cleaned test split, where predicted-versus-experimental scatter retains a tight envelope. A simple inductive bias further enhances generalization across devices of differing initial quality. The baseline captures systematic device-level effects, allowing the model to focus capacity on R(X), the chemical contribution; this reduces error and improves robustness in prospective screening.

**Model Hypotheses:**

Hypotheses 1: (Linear Trend Prior): We hypothesize that the expected ΔPCE, averaged over the chemical space of additives, exhibits a negative linear dependency on the initial efficiency I. This baseline trend is captured by a linear regression model fitted exclusively on the training and validation data to prevent data leakage.

$$E[\Delta\mathrm{PCE}|I] = kI + b.$$

We define this based on the training and validation set as the baseline enhancement:

$$\mathrm{Linear}_{\Delta\mathrm{PCE}}(I) = kI + b$$

Hypotheses 2: (Residual Decomposition): We posit that the actual ΔPCE for a specific molecule X and I can be decomposed into the linear baseline and a molecule-specific residual term.

$$\Delta\mathrm{PCE}(X, I) = \mathrm{Linear}_{\Delta\mathrm{PCE}} + R(X).$$

**QSPR Learning Objective:**

The central goal of our machine learning model is not to predict ΔPCE directly, but rather to learn R(X), which represents the molecule's ability to perform better or worse than the statistical average for a given I. The model, $f_{QSPR}$ learns the mapping:

$$f_{QSPR}: X \to \hat{R}(X) \approx R(X)$$

For each data point $(X_i, I_i, \Delta PCE_i)$ in the training set, the target label for supervised learning is computed as the true residual:

$$R_i \coloneqq \Delta PCE_i - (kI_i + b)$$

**Inference and Final Prediction:**

For a new, unseen molecule $X^*$ tested on a device with a known initial efficiency $I^*$, the final predicted efficiency enhancement is synthesized by combining the linear baseline with the model's predicted residual:

$$\Delta\text{PCE}_{pred}(X^*, I^*) = (kI^* + b) + \hat{R}(X^*) = \text{Linear}_{\Delta\text{PCE}}(I^*) + f_{\text{QSPR}}(X^*)$$

This composite approach allows the model to make a more nuanced prediction that accounts for both the initial state of the system and the unique chemical properties of the additive, a concept central to advanced molecular regression techniques.

**Experimental Justification and Performance Impact**

The validity of incorporating this inductive bias is strongly supported by the fundamental photophysics of solar cells. The negative correlation between Initial_PCE and ΔPCE is not merely a statistical artifact but is physically grounded in the Shockley-Queisser (SQ) limit, which defines the theoretical maximum efficiency for a given semiconductor bandgap. A device with a higher initial PCE is already operating closer to this physical ceiling. Consequently, there is diminished headroom for further improvement, and any additive, no matter how effective, will produce a smaller absolute ΔPCE. Our framework correctly captures this "diminishing returns" phenomenon by treating it as a systemic baseline, thereby allowing the QSPR model to focus on identifying the intrinsic passivating or performance-enhancing properties of the molecule itself.

By adopting this inductive bias-informed framework, we achieve a marked improvement in predictive accuracy. A naive QSPR model that attempts to predict ΔPCE directly from molecular features X is forced to implicitly learn both the complex molecular effects and the simple, dominant linear trend associated with I. This conflation of signals can lead to a less robust model with lower predictive power. By explicitly decoupling these two components, our method refines the learning task. The QSPR model is no longer burdened with rediscovering the I-dependent trend and can instead dedicate its full capacity to modeling the residual R(X), a target variable with lower variance. This results in a model with a significantly higher $R^2$ and lower overall error when compared to the direct prediction approach, demonstrating a more accurate and generalizable understanding of the structure-property relationship governing ΔPCE enhancement in perovskite solar cells

**Supplementary Note 3 Feature visualization and interpretation**

For feature visualization, we extracted the representations from the final layer of the model's classification head and projected them into two-dimensional space using UMAP. The UMAP hyperparameters were configured as follows:

| Hyperparameter | Definition | Value/Setting |
|---|---|---|
| n_components | Output dimensionality | 2 |
| random_state | Random seed for reproducibility | 42 |
| n_neighbors | Local neighborhood size | 15 |
| min_dist | Minimum embedded point spacing | 0.1 |
| metric | Distance metric | 'euclidean' |
| n_epochs | Optimization iterations | 200 |

For atomic heatmap generation, we computed the attention mappings from the last layer of the model's self-attention mechanism. The attention weights were averaged across all attention heads of the last attention layer and subsequently normalized to produce the final interpretable heatmaps.

To generate molecule features we used Rdkit[3], Gaussian 16, CP2K[4] Multiwfn[5], VESTA[6], Pymatgen[7].

## Supplementary Note 4: Perovskite solar cells experimental methods

### Materials and reagents.

All chemicals and materials were employed without further purification except where noted. Commercial sources supplied the reagents as follows: Advanced Election Technology provided patterned fluorine-doped tin oxide substrates (FTO, AGC, 7 $\Omega$ $sq^{-1}$, 2.2 mm thick). Formamidinium iodide (FAI, >99.99%), methylammonium chloride (MACl, 99.9%), lead iodide ($PbI_2$, 99.99%) and cyclohexanemethylamine hydroiodide (CHMAI, 99.9%) were acquired from Tokyo Chemical Industry Co., Ltd. (TCI). Mesoporous $TiO_2$ paste (30 NR-D), produced by Greatcell Solar Materials, was supplied by Yingkou Opvtech New Energy. Sigma-Aldrich supplied titanium diisopropoxide bis(acetylacetonate) (75% in isopropanol) and acetylacetonate (⩾ 99%). The 2,2',7,7'-tetrakis(N,N-di-p-methoxyphenyl-amine)-9,9'-spirobifluorene (Spiro-OMeTAD, 99.9%) was acquired from Borun New Material Technology. Additional additives included lithium bis(trifluoromethylsulfonyl)imide (LiTFSI, 99.95%) and 4-tert-butylpyridine (tBP, 96%). Dimethylformamide (DMF, 99.8%), dimethyl sulfoxide (DMSO, 99.7%), isopropanol (IPA, 99.5%), ethanol (EtOH, 99.5%), 2-methoxyethanol (2-MeOEtOH, 99.8%), acetonitrile (ACN, 99.9%), and chlorobenzene (CB, 99.8%) were procured from Acros Organics. A 0.22 μm polytetrafluoroethylene filter was used to filter all solvents prior to utilization. Sinopharm Chemical Reagent provided EtOH, acetone and deionized (DI) water for cleaning purposes. Gold (Au) with a purity of ≥99.99% was obtained from commercial suppliers. $FAPbI_3$ powder was synthesized in-house by reacting FAI and $PbI_2$ in 2-methoxyethanol, following a published method[8].

### Perovskite precursor solution.

The compact $TiO_2$ (c-$TiO_2$) solution was formulated by diluting 0.6 mL titanium diisopropoxide bis(acetylacetonate) and 0.4 mL acetylacetonate with 9 mL anhydrous ethanol. The diluted mesoporous $TiO_2$ (mp-$TiO_2$) paste was obtained by mixing $TiO_2$ paste (30 NR-D) with ethanol at a weight ratio of 1:6. The perovskite precursor was prepared by dissolving 1.8 M α-phase $FAPbI_3$ powder and 0.64 M MACl in 1 mL of a DMF:DMSO mixed solvent (v:v, 86:14). The DFBP-treated and *o*-TCPN-treated perovskite precursor was prepared by adding DFBP and *o*-TCPN to the control precursor at concentrations of 0.1 mg $mL^{-1}$ and 0.15 mg $mL^{-1}$, respectively. For the hole transport layer (HTL), the HTL precursor comprises 72.3 mg Spiro-OMeTAD, 28.8 μL tBP, and 17.5 μL Li-TFSI (520 mg $mL^{-1}$ in ACN), all dissolved in 1 mL CB. The post-treatment solution is composed of 6 mg $mL^{-1}$ CHMAI dissolved in IPA.

**Device fabrication.**

The perovskite solar cells (PSCs) were constructed with a device structure of glass/FTO/c-$TiO_2$/mp-$TiO_2$/perovskite/Spiro-OMeTAD/Au. The glass/FTO substrates were thoroughly cleaned using a sequence of soapy water, deionized water, acetone, and ethanol， with each cleaning step lasting for 15 minutes. The c-$TiO_2$ layer was deposited on the cleaned glass/FTO substrates via spray pyrolysis at 450°C, using oxygen as the carrier gas. The mp-$TiO_2$ layer was formed by spin-coating the diluted paste at 4000 rpm, followed by sintering at 450°C for 30 minutes in dry air. Then, the substrates were treated with $Li_2CO_3$ aqueous solution and sintered at 450 °C for 30 min in ambient air again. After cooling to room temperature, the substrates were treated with UV-Ozone for 10 minutes and then transferred into a nitrogen glovebox. The perovskite precursor was spin-coated onto the substrate using a two-step program: initially spinning at 2000 rpm with a ramping rate of 200 rpm $s^{-1}$ for 10 seconds, followed by 5000 rpm with a ramping rate of 2000 rpm $s^{-1}$ for 22 seconds. At 25 seconds, 100 μL of the antisolvent CB was dropped onto the substrate surface. The wet film was then transferred out of the glovebox and thermally annealed at 150°C in ambient air with a relative humidity (RH) of 30-40%. Subsequently, the perovskite-coated substrate was returned to the nitrogen glovebox and annealed at 100°C for 15 minutes. After cooling, the post-treatment solution was deposited onto the perovskite-coated substrate by spin-coating at 6000 rpm for 30 seconds with a ramping rate of 3000 rpm $s^{-1}$, followed by thermal annealing at 105°C for 5 minutes. The HTL precursor was then spin-coated at 5000 rpm for 30 seconds with a ramping rate of 3000 rpm $s^{-1}$. Finally, an 80 nm-thick gold layer was thermally evaporated onto the films in a high vacuum chamber ($<6\times10^{-4}$ Pa).

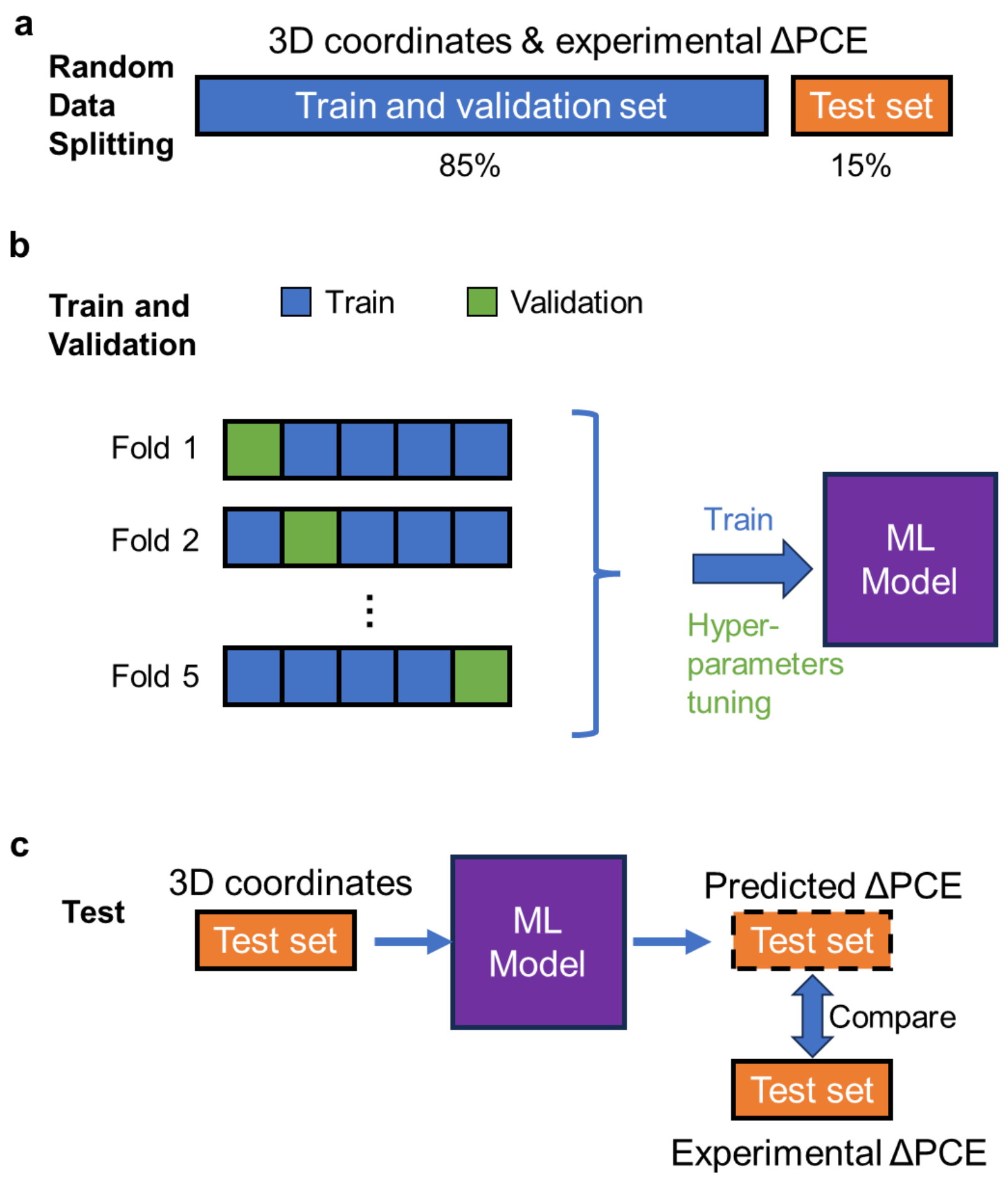

**Supplementary Figure 1 |** Three-stage benchmark workflow for ML QSAR model training and evaluation. (a) Data partitioning into training/validation sets (via cross-validation) and an independent test set. (b) Hyper-parameter tuning optimization using training data while maintaining test data separation. (c) Final model evaluation on the test set for unbiased performance assessment. This workflow minimizes overfitting through proper data separation and ensures reproducible validation. For a given dataset and a given seed, we partitioned it into test sets and train & validation sets at an 85:15 ratio.

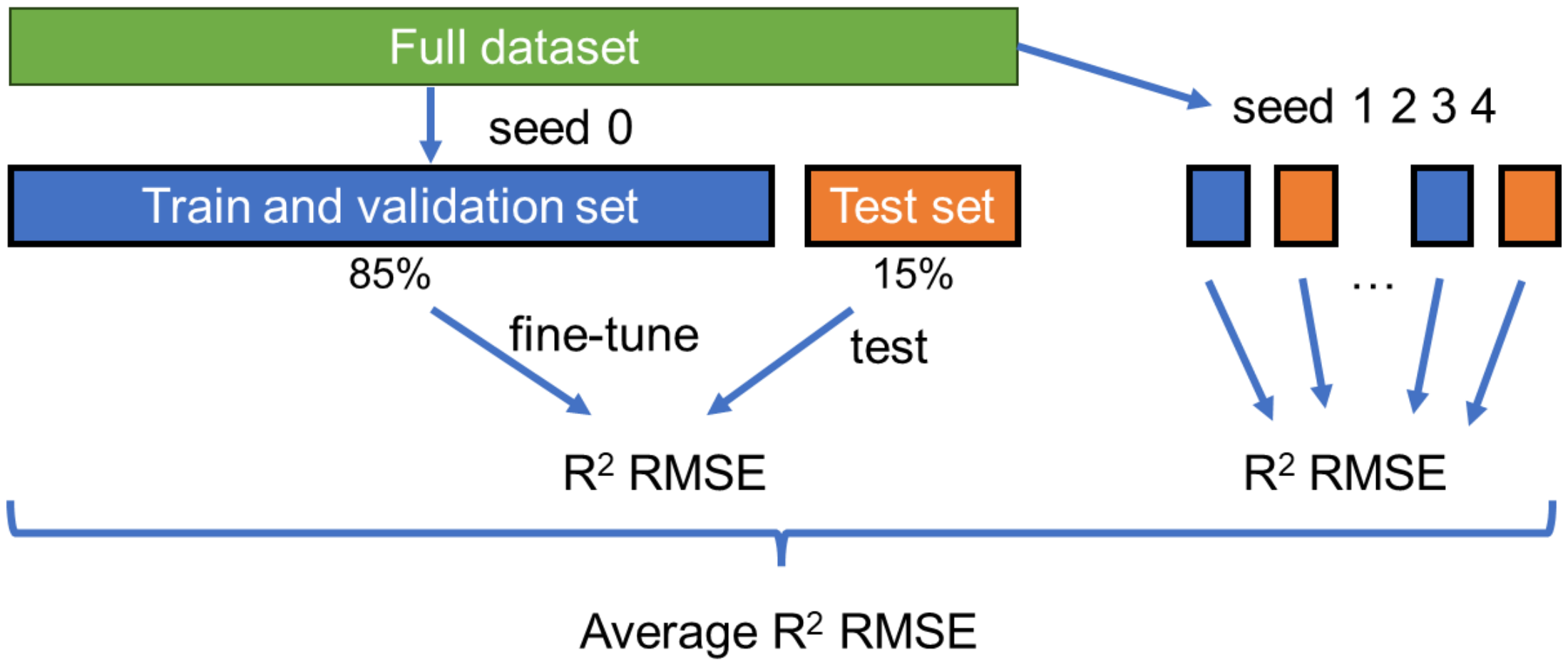

**Supplementary Figure 2 |** Training and testing were repeated across 5 independent random seeds (0 to 4) to report the average $R^2$ and RMSE to assess model robustness of the ML model.

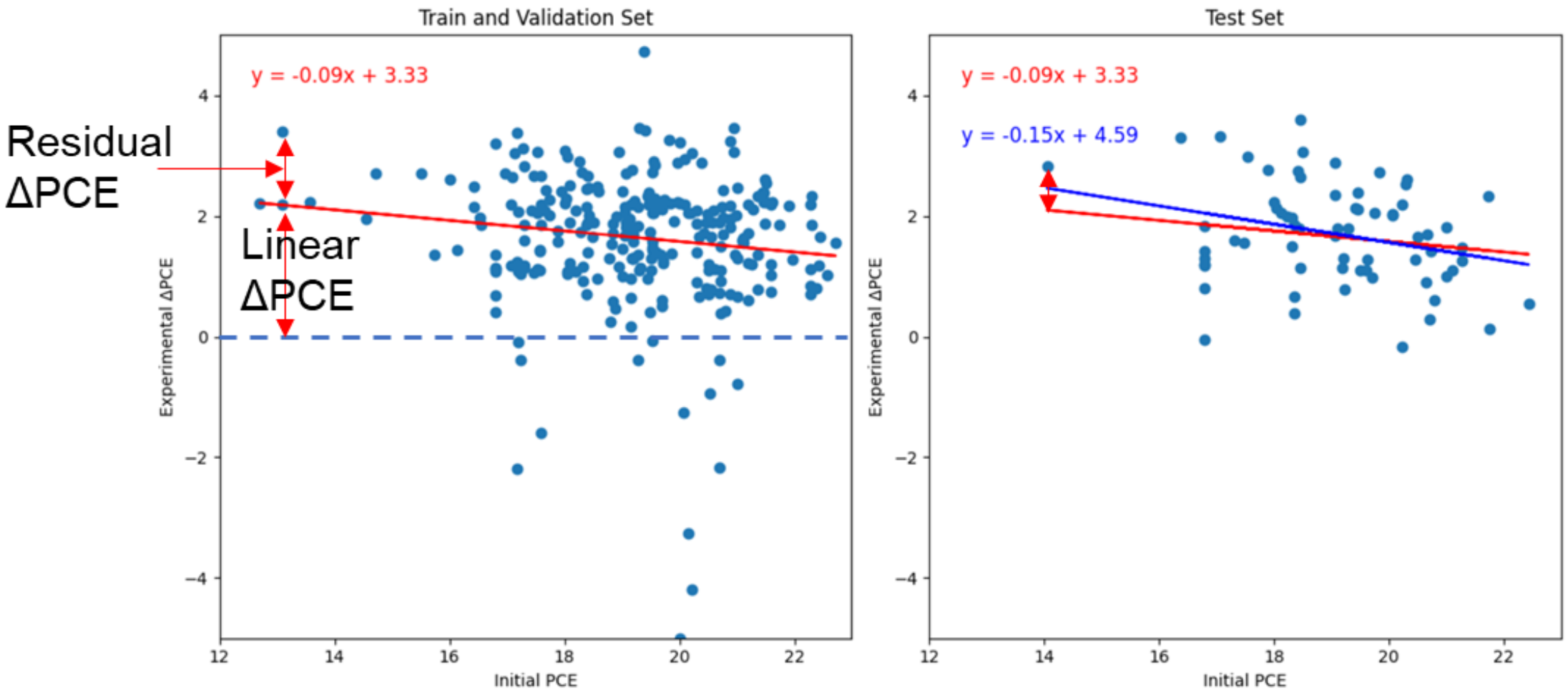

**Supplementary Figure 3 |** Inductive bias separating baseline and residual separation. A linear baseline relates initial PCE to expected ΔPCE on an average molecule, while the ML model learns molecule-specific residuals $R$(X). For the given dataset split, only the baseline equation $y$ = -0.09$x$ + 3.33 was used for model training and testing. However, the linear correlation $y$ = -0.15$x$ + 4.59 can be calculated from the test set, which is similar to the train and validation set calculated baseline, confirming the generalization ability of this method.

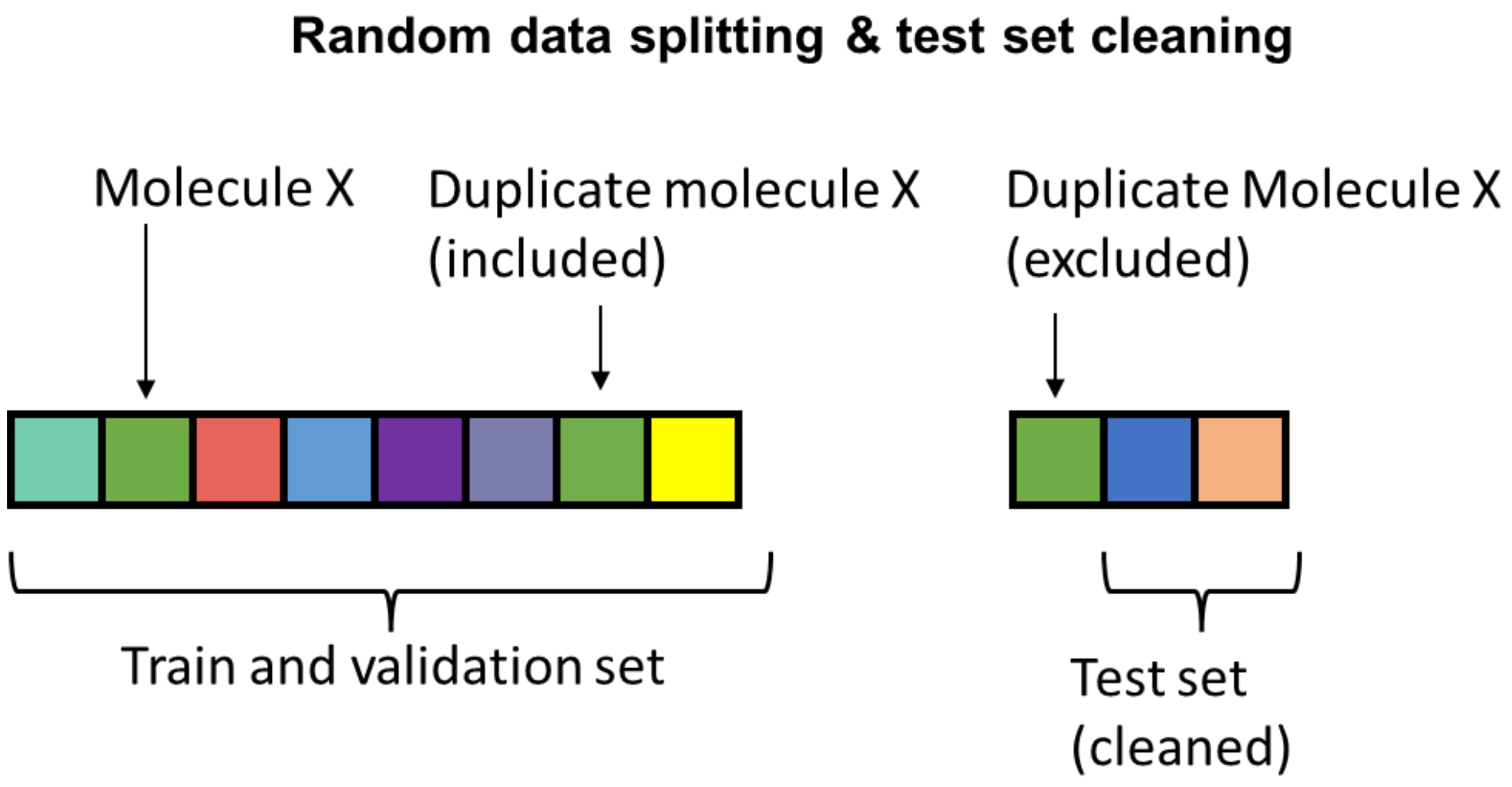

**Supplementary Figure 4 |** Duplicate data cleaning to prevent data leakage when testing. Some molecules were reported several times by different literatures (often with different $\Delta PCE_{exp}$ on different compositions of perovskite in PSCs), which could potentially enrich our dataset and work as data augmentation, so we preserve the duplicate molecules in the train and validation set. However, when a molecule appears in both the train and validation set, the one in the test set is removed to prevent data leakage.

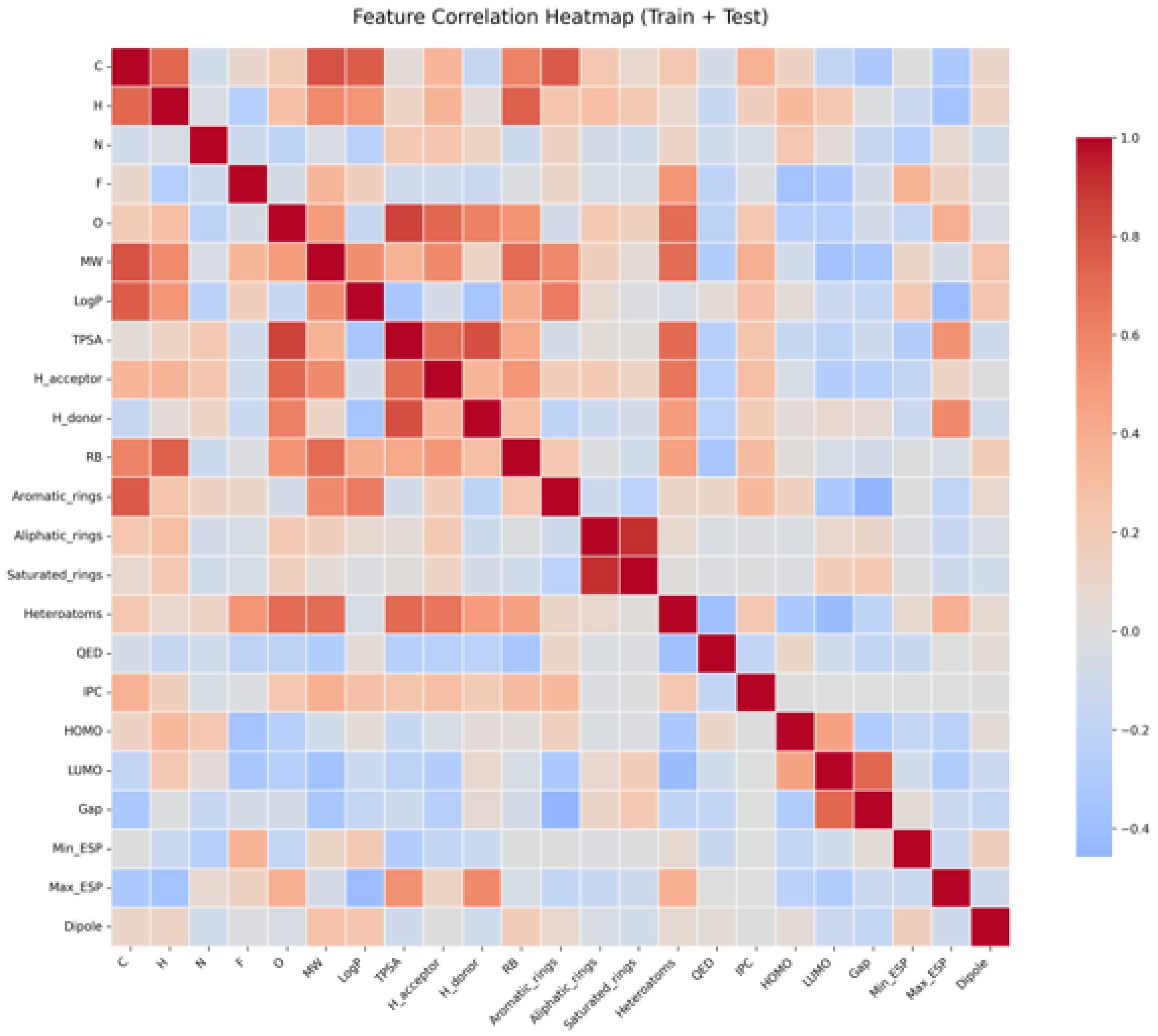

**Supplementary Figure 5 |** Feature correlation heatmap for conventional descriptor sets. Highly correlated feature pairs are shown in red and will be considered for removal to reduce multicollinearity among features and prevent overfitting.

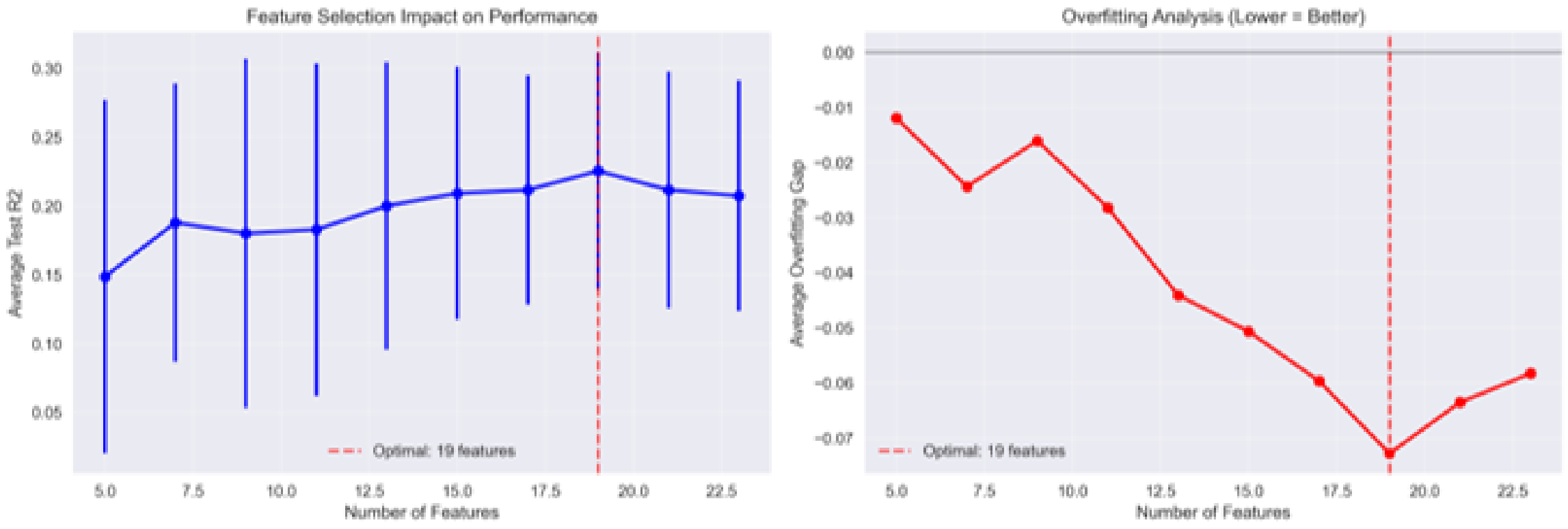

**Supplementary Figure 6 |** In the DFT-feature-based ML paradigm, models with too few features tend to underfit, whereas models with an excessive number of features are prone to overfitting.

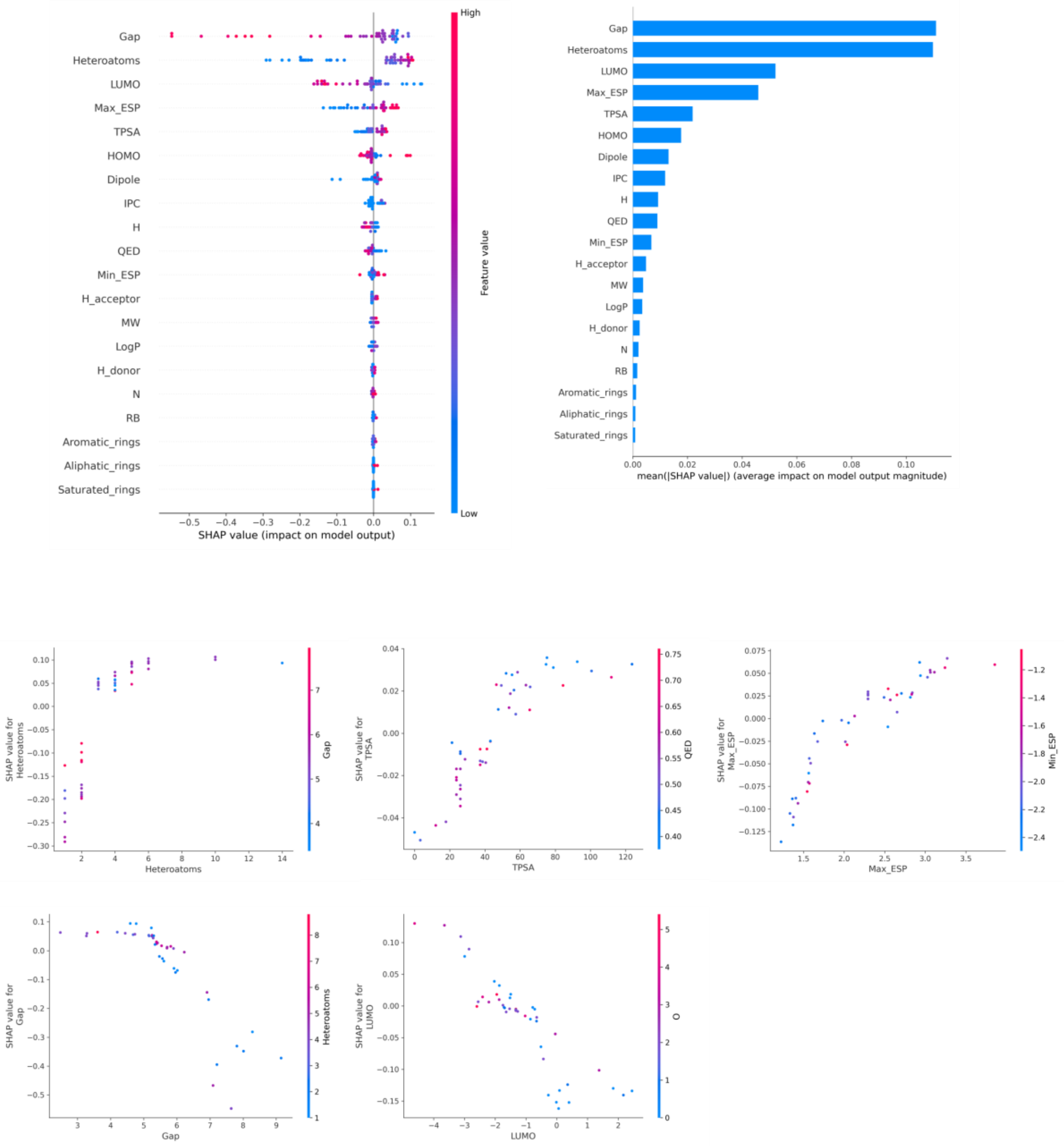

**Supplementary Figure 7 |** SHAP analysis for DFT-calculated descriptors to quantify and rank their contributions to the predicted ΔPCE.

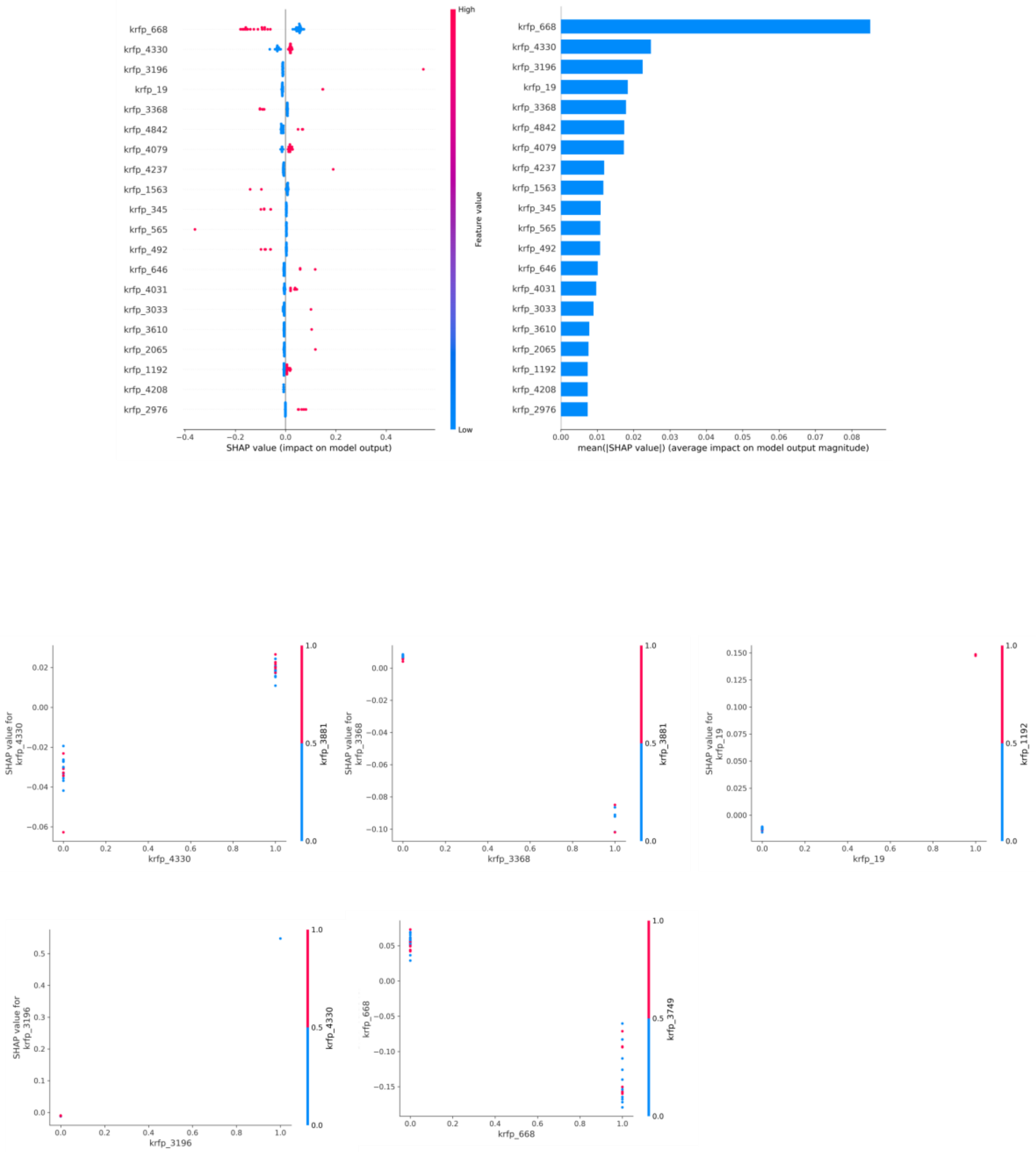

**Supplementary Figure 8 |** SHAP analysis for KRFP fingerprints, highlighting the molecular substructures most strongly associated with ΔPCE.

Top 100 Molecules by Prediction Value

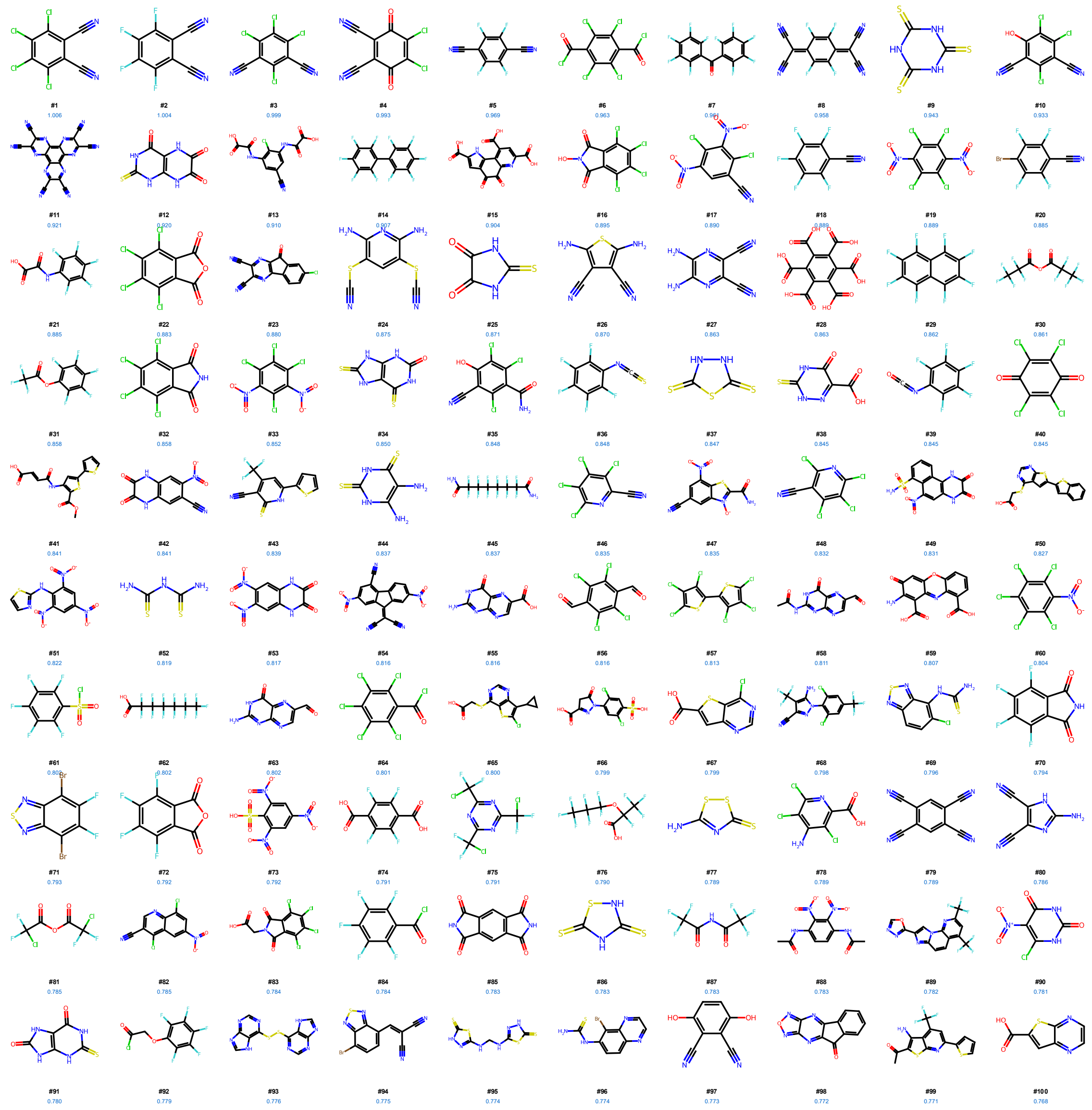

**Supplementary Figure 9 |** Top-100 molecules predicted by Uni-Mol model. Displays the highest-ranked candidates based on predicted ΔPCE.

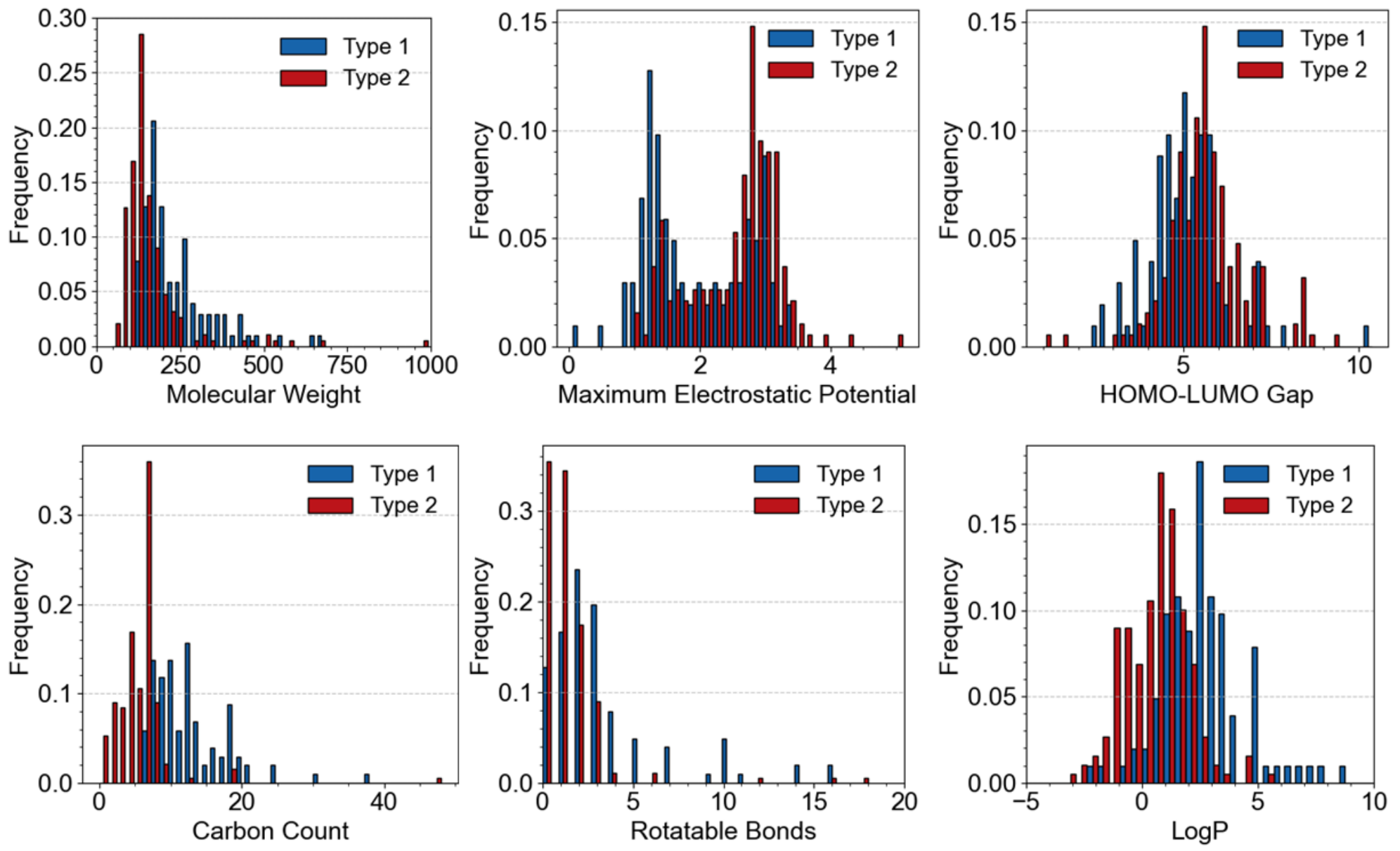

**Supplementary Figure 10 |** DFT-calculated features exhibiting significant differences between Type 1 and Type 2 molecular clusters. These properties distinguish across clusters, and contribute to the primary variation of the Uni-Mol representation captured chemical features.

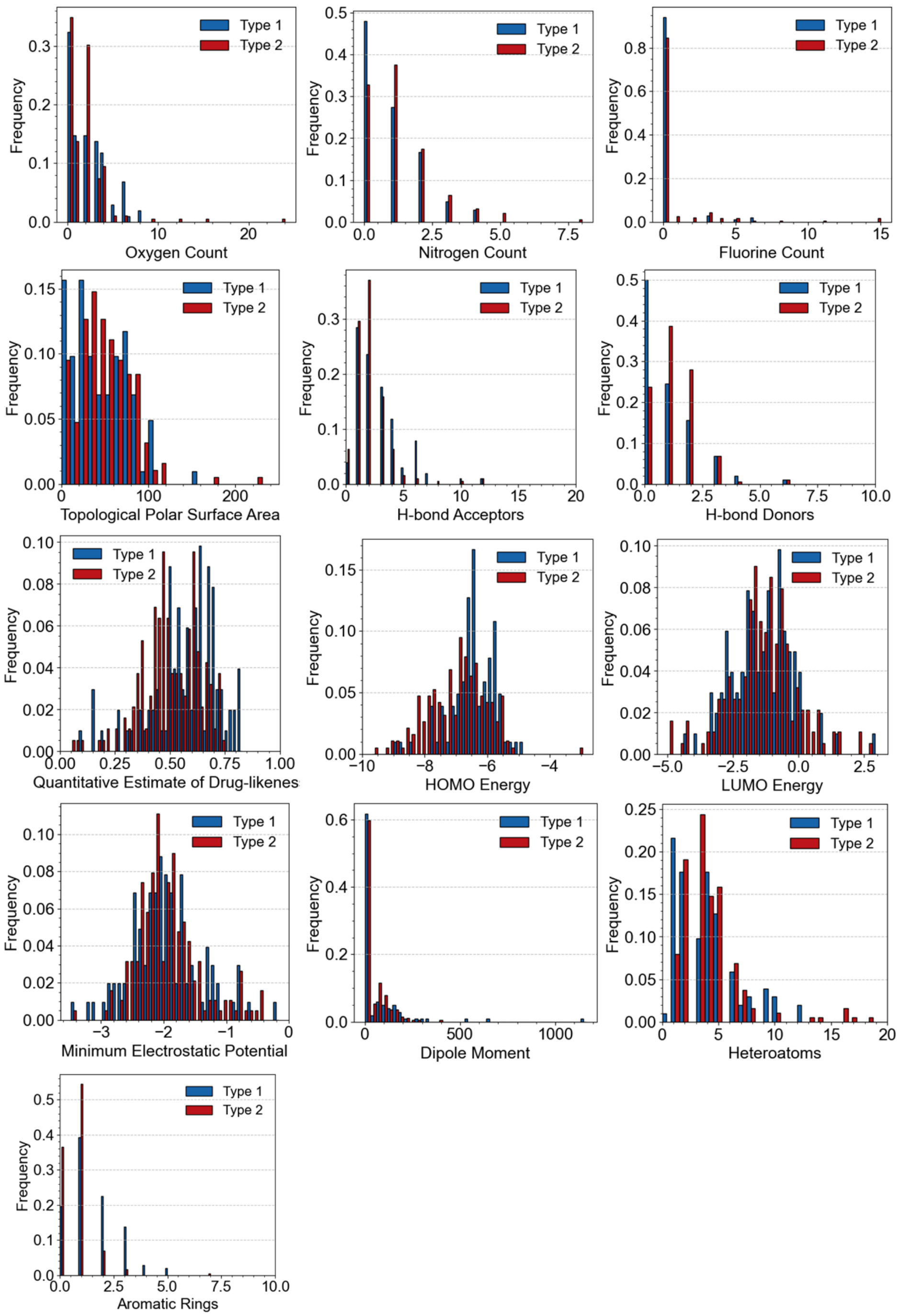

**Supplementary Figure 11 |** DFT-calculated features showing minimal differences between Type 1 and Type 2 molecular clusters. These properties remain largely consistent across clusters, and do not contribute to the primary variation of the Uni-Mol representation captured chemical features.

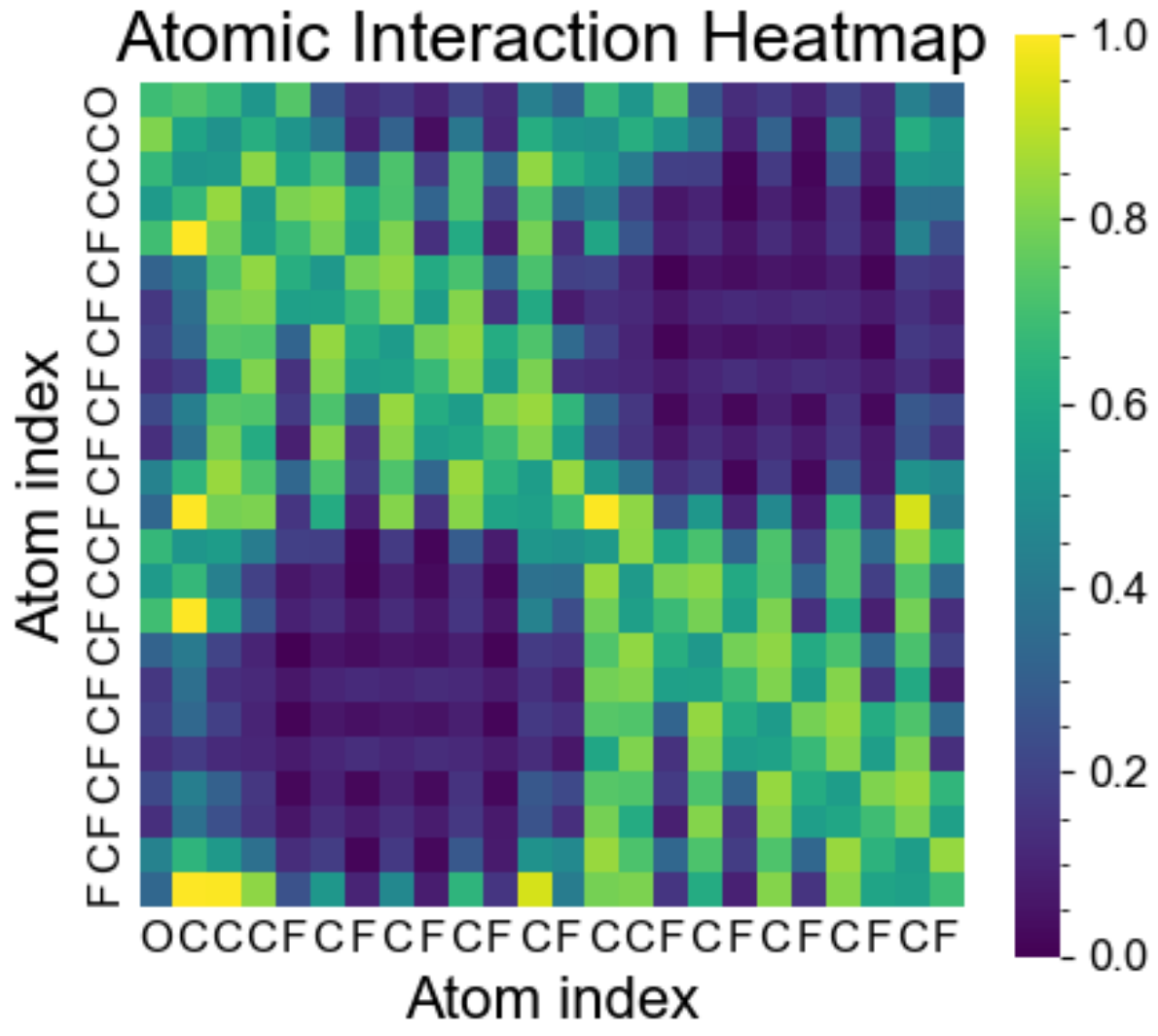

**Supplementary Figure 12 |** The attention heatmap of DFBP. Highlights the atomic and bond-level contributions identified by the fine-tuned Uni-Mol model, emphasizing functional groups and structural motifs most relevant for interaction with the perovskite and modulation of ΔPCE.

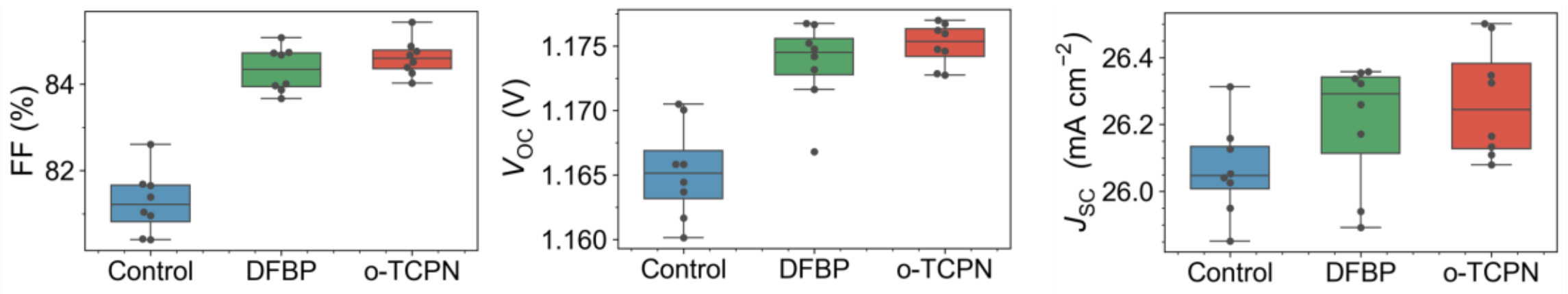

**Supplementary Figure 13 |** Statistics of device parameters including current density ($J_{SC}$), open-circuit voltage ($V_{OC}$), fill factor (FF) across 8 PSC devices for each evaluated molecule. The center line represents the median, the shaded box spans the 25$^{th}$-75th percentiles, and whiskers extend to 1.5× the interquartile range.

**Supplementary Table 1 |** Fine-tuning comparison of pre-trained versus randomly initialized backbones.

| | RMSE | $R^2$ |
|---|---|---|
| pre-trained backbone | **0.486** | **0.502** |
| random backbone | 0.545 | 0.375 |

**Supplementary Table 2 |** Molecular properties used for traditional ML.

| Feature | Calculation_Method | Dimension | Feature_Definition | Detailed_Interpretation |
|---|---|---|---|---|
| C | RDKit | 1D | Carbon Atom Count | Number of carbon atoms in the molecule. |
| H | RDKit | 1D | Hydrogen Atom Count | Number of hydrogen atoms in the molecule. |
| N | RDKit | 1D | Nitrogen Atom Count | Number of nitrogen atoms in the molecule. |
| F | RDKit | 1D | Fluorine Atom Count | Number of fluorine atoms in the molecule. |
| O | RDKit | 1D | Oxygen Atom Count | Number of oxygen atoms in the molecule. |
| MW | RDKit | 1D | Molecular Weight | The mass of the molecule, calculated as the sum of atomic weights. |
| LogP | RDKit | 2D | Octanol-Water Partition Coefficient | A measure of the molecule's lipophilicity or hydrophobicity. Higher value indicates greater fat solubility. |
| TPSA | RDKit | 2D | Topological Polar Surface Area | The sum of the surface areas of polar atoms (typically oxygen, nitrogen, and attached hydrogens). Related to drug transport properties. |
| H_acceptor | RDKit | 2D | Hydrogen Bond Acceptors | The count of atoms (usually N or O) capable of accepting a hydrogen bond. |
| H_donor | RDKit | 2D | Hydrogen Bond Donors | The count of atoms (usually N-H or O-H) capable of donating a hydrogen bond. |
| RB | RDKit | 2D | Rotatable Bonds | The number of non-terminal single bonds whose rotation is not constrained by a ring. Measures molecular flexibility. |
| Aromatic_rings | RDKit | 2D | Aromatic Ring Count | The total number of rings that exhibit aromaticity. |
| Aliphatic_rings | RDKit | 2D | Aliphatic Ring Count | The total number of non-aromatic rings (includes saturated and unsaturated non-aromatic rings). |
| Saturated_rings | RDKit | 2D | Saturated Ring Count | The total number of rings where all constituent atoms are saturated (connected by only single bonds within the ring). |
| Heteroatoms | RDKit | 2D | Heteroatom Count | The total number of non-carbon and non-hydrogen atoms in the molecule. |
| QED | RDKit | 2D | Quantitative Estimate of Drug-likeness | A composite score that assesses the likelihood of a molecule being a drug candidate based on multiple properties. |
| IPC | RDKit | 2D | Information Content Index | A topological descriptor quantifying the structural complexity and degree of asymmetry of the molecule. |
| HOMO | DFT | 3D | Highest Occupied Molecular Orbital energy | The energy of the highest electron orbital that is occupied. Relates to the molecule's electron-donating ability. |
| LUMO | DFT | 3D | Lowest Unoccupied Molecular Orbital energy | The energy of the lowest electron orbital that is unoccupied. Relates to the molecule's electron-accepting ability. |
| Gap | DFT | 3D | HOMO-LUMO Energy Gap | The energy difference (LUMO - HOMO). Predicts chemical reactivity and kinetic stability; a smaller gap suggests higher reactivity. |
| Min_ESP | DFT | 3D | Minimum Electrostatic Potential | The most negative electrostatic potential found on the molecular surface. Indicates the most favorable site for electrophilic attack. |
| Max_ESP | DFT | 3D | Maximum Electrostatic Potential | The most positive electrostatic potential found on the molecular surface. Indicates the most favorable site for nucleophilic attack. |
| Dipole | DFT | 3D | Total Dipole Moment | The magnitude of the net molecular polarity. Represents the overall separation of positive and negative charges in the molecule. |

**Supplementary Table 3 |** Traditional Machine learning algorithms.

| ML Algorithm | Description |
|---|---|
| Linear Regression | A simple linear model that models the relationship between the features and the target variable by fitting a linear equation to the observed data. |
| Ridge Regression | A linear model regularized with Tikhonov regularization of L2 penalty to prevent overfitting by shrinking the coefficients. |
| Lasso Regression | A linear model regularized with an L1 penalty, which promotes sparsity by driving some coefficients exactly to zero. |
| Elastic Net | A linear model regularized with both L1 and L2 penalties, balancing sparsity and coefficient shrinkage. |
| Bayesian Ridge | A probabilistic linear regression model that estimates a posterior distribution over the coefficients. |
| Decision Tree Regressor | A non-linear model that partitions the feature space into a set of rectangular regions to make predictions. |
| Random Forest Regressor | An ensemble method combining multiple decision trees with bootstrap aggregating to improve prediction accuracy and reduce overfitting. |
| Gradient Boosting Regressor | An ensemble method that sequentially builds a strong prediction model from a series of weak prediction models using gradient descent. |
| Support Vector Regressor | A model that uses the kernel trick to map data to a higher dimension for non-linear regression, aiming to fit the data within an epsilon-margin. |
| K-Nearest Neighbors Regressor | A non-parametric, instance-based learning method that predicts the target value based on the average of the target values of its k nearest neighbors. |
| Multi-layer Perceptron Regressor | A feedforward artificial neural network model with one or more hidden layers. |
| Gaussian Process Regressor | A non-parametric, probabilistic model that defines a prior over functions, providing uncertainty estimates along with predictions. |
| Kernel Ridge Regression (RBF Network) | A regularized linear model in a kernel-induced feature space. |
| XGBoost Regressor | An optimized distributed gradient boosting library designed to be highly efficient, flexible, and portable. |
| LightGBM Regressor | A fast, distributed, high-performance gradient boosting framework that uses tree-based learning algorithms, notable for its use of Gradient-based One-Side Sampling (GOSS). |
| CatBoost Regressor | An open-source gradient boosting library that utilizes symmetric decision trees and is notable for its excellent handling of categorical features and lack of overfitting. |

**Supplementary Table 4 |** Benchmark of different machine learning paradigms performance on the perovskite dataset.

| Paradigm | Model | $R^2$ (avg±std) | RMSE (avg±std) |
|---|---|---|---|
| RDKit+DFT | Bayesian Ridge | 0.149 ± 0.154 | 0.638 ± 0.064 |
| | Decision Tree Regressor | 0.068 ± 0.201 | 0.667 ± 0.074 |
| | Elastic Net | 0.141 ± 0.152 | 0.642 ± 0.069 |
| | Gaussian Process Regressor | -2.946 ± 6.410 | 1.124 ± 0.897 |
| | Gradient Boosting Regressor | 0.184 ± 0.161 | 0.624 ± 0.064 |
| | K-Nearest Neighbors Regressor | 0.156 ± 0.091 | 0.638 ± 0.059 |
| | Lasso Regression | 0.141 ± 0.159 | 0.642 ± 0.073 |
| | Linear Regression | 0.033 ± 0.208 | 0.678 ± 0.059 |
| | Multi-layer Perceptron Regressor | 0.063 ± 0.170 | 0.669 ± 0.049 |
| | Kernel Ridge Regression (RBF Network) | 0.121 ± 0.188 | 0.647 ± 0.066 |
| | **Random Forest Regressor** | **0.199 ± 0.135** | **0.620 ± 0.069** |
| | Ridge Regression | 0.143 ± 0.153 | 0.641 ± 0.066 |
| | Support Vector Regressor | 0.185 ± 0.115 | 0.626 ± 0.058 |
| KRFP | Bayesian Ridge | 0.112 ± 0.093 | 0.656 ± 0.080 |
| | Decision Tree Regressor | -0.005 ± 0.151 | 0.696 ± 0.081 |
| | Elastic Net | 0.108 ± 0.119 | 0.657 ± 0.086 |
| | Gaussian Process Regressor | -0.058 ± 0.145 | 0.718 ± 0.117 |
| | **Gradient Boosting Regressor** | **0.127 ± 0.100** | **0.652 ± 0.094** |
| | K-Nearest Neighbors Regressor | -0.081 ± 0.190 | 0.720 ± 0.079 |
| | Lasso Regression | 0.068 ± 0.112 | 0.672 ± 0.085 |
| | Linear Regression | -1.150 ± 0.363 | 1.015 ± 0.097 |
| | Multi-layer Perceptron Regressor | -0.270 ± 0.307 | 0.776 ± 0.061 |
| | Kernel Ridge Regression (RBF Network) | 0.077 ± 0.137 | 0.667 ± 0.077 |
| | Random Forest Regressor | 0.117 ± 0.117 | 0.656 ± 0.104 |
| | Ridge Regression | 0.104 ± 0.123 | 0.657 ± 0.068 |
| | Support Vector Regressor | 0.061 ± 0.149 | 0.672 ± 0.075 |
| **Transfer Learning** | ChemBERTa2 | 0.2025 ± 0.0781 | 0.6327 ± 0.0531 |
| | MolCLR | 0.1725 ± 0.1095 | 0.6465 ± 0.0743 |
| | **Uni-Mol** | **0.3208 ± 0.1088** | **0.5764 ± 0.0708** |

**Supplementary Table 5 |** Summary of different machine learning QSPR models from the literature.

| Paper_DOI | Input_Molecule_Properties | Input_Perovskite_Properties | Trainset_Size | Testset_Size | Perovskite_Device_Type | Molecule_Type | ML_Algorithm | Output_Properties |
|---|---|---|---|---|---|---|---|---|
| 10.1002/adfm.202314529 | 1D properties+3D quantum properties | | 310 | | n-i-p solar cells | bulk doping | SVM NNM RF KNN NB S-R model (SVM+RF) Ensemble | PCE(modified)-PCE(unmodified) |
| 10.48550/arXiv.2412.14109 | MACCS fingerprints RDKit molecular descriptors JTVAE-latent vectors | | 129 | | p-i-n solar cells | bulk doping+posttreatment | RF GB SVM | PCE(modified) |
| 10.1126/science.ads0901 | Substituent species, Aromatic ring species, Conjugate length, Flexible and rigid units, Electronic effect, Spatial effect, HOMO, LUMO, Rotation constant b, Number of atoms, Molecular weight, Molecular LogP | | 149 | | p-i-n solar cells | HTM | Gaussian Process, initially used as a surrogate model for Bayesian Optimization, and later combined with a kernel machine for a Recursive Feature Machine (RFM) approach. | PCE(modified)FFJscVoc |
| 10.1021/acsenergylett.4c02610 | Electrostatic potential, energy levels, molecular polarity, molecular descriptors (e.g., topological polar surface area, molecular fingerprints) | | 175 | 15% of the dataset (approximately 26) | n-i-p solar cells | SAMs | XGBoost, Deep Neural Network (DNN), Chemprop (directed message-passing neural network) | Power Conversion Efficiency (PCE), PCE increase, PCE increase percentage |
| 10.1016/j.cej.2024.156391 | Molecular complexity, molecular weight, number of O atoms, hydrogen bond acceptors, molecular descriptors (e.g., topological polar surface area, molecular fingerprints) | $MAPbI_3$-based perovskite solar cells (PSCs) | 63 | Approximately 20% of the dataset (around 13) | $MAPbI_3$-based perovskite solar cells | Organic small molecules (additives) | Random Forest (RF), Gradient Boosting (GB), Extreme Gradient Boosting (XGBoost), Support Vector Regression (SVR), Linear Regression (LR) | Power Conversion Efficiency (PCE) improvement rate |
| 10.1021/acs.jctc.4c00465 | Textual descriptors (from NLP model) + DFT descriptors (total dipole moments, energy gap, HOMO, LUMO) | Aqueous photocurrents of multisolvent engineered halide perovskite | | | | Solvent molecules | Genetic Algorithm (GA) | Aqueous photocurrent stability |
| | | CH3NH3PbI3 | | | | | | |
| 10.1021/acsami.4c06226 | Small molecule descriptors (Eg, Size, Rn, Trap, TRPL) | Performance data of inorganic perovskite materials | 86 | 22 | All-inorganic perovskite solar cells | Organic small molecules | Random Forest (RF), Support Vector Machine (SVR), XGBoost | PCE, JSC, VOC, FF |
| 10.26434/chemrxiv-2024-b9rw6 | 42 molecular features (e.g., avgpol, axxpol, ayypol, azzpol, molpol, ASA+, ASA-, ASA_H, ASA_P, asa, etc.) | CsPbBr3 perovskite nanocrystals | 21 (initial) + 72 (model-recommended) | | | Organic small molecules | Twin regressor with random forest base predictor | Relative PL enhancement (FOM) |
| 10.1002/adts.202300978 | 12 features (e.g., logM, Balaban, tpsa, logP, Kappa, bertz_ct, HOMO, chi1, Wiener, MolWt, HBO, HBA) | | 108 | 33 | n-i-p solar cells | HTMs | Random Forest | logM (logarithm of Hole Mobility) |
| 10.1002/adfm.202314529 | 31 features (e.g., dipole moment, HOMO-LUMO gap, molar volume, rotatable bond count, etc.) | | 330 | | n-i-p solar cells | Small molecules | Random Forest, Neural Network Model, Support Vector Regression, Decision Tree, Naive Bayes | PCE (Power Conversion Efficiency) |
| 10.1039/d4nj03777d | 1D, 2D, 3D molecular descriptors, 2D binary fingerprints | | 124 | | n-i-p and p-i-n solar cells | HTMs | Kernel Partial Least Square (KPLS), Partial Least Square (PLS), Principal Component Regression (PCR), Multiple Linear Regression (MLR) | PCE, VOC, JSC |
| 10.1039/d4ta03547j | 3D-MoRSE descriptors (morU, morE, morIP) | | 105 | | n-i-p solar cells | Passivation molecules | Automatic Relevance Determination Regression | PCE (Power Conversion Efficiency) |
| 10.1038/s41563-023-01705-y | 19 features (e.g., electronic parameters, structural parameters, fundamental parameters) | | 168 | | Inverted p-i-n solar cells | Pseudo-halide (PH) anions | Random Forest, Logistic Regression | Eb (binding energy) classification |
| 10.1002/solr.202300490 | Various features (e.g., perovskite composition, annealing temperature, HTL/ETL properties, etc.) | | Over 40,000 | | Various solar cells | HTL and ETL materials, perovskite additives | Gradient-Boosted Regression Trees (LightGBM) | PCE (Power Conversion Efficiency) |
| 10.1016/j.jechem.2023.04.015 | Electron topological-state (E-state) index, cheminformatics features, etc. | Ion stoichiometry, HTL type | 303 | | p-i-n type solar cells | Interface materials | Random Forest, Linear Regression, Extreme Gradient Boosting, Neural Network | VOC (Open-Circuit Voltage) |
| **This work** | **Uni-Mol representations** | **FAPbI3-based PSCs** | **343*85%** | **343*15%** | **n-i-p type solar cells** | **bulk modulator molecules** | **Transformers encoder** | **ΔPCE** |

## References

1. Zhou, G. *et al.* Uni-Mol: A Universal 3D Molecular Representation Learning Framework. in (ICLR, 2023). doi:10.26434/chemrxiv-2022-jjm0j-v4.
2. Akiba, T., Sano, S., Yanase, T., Ohta, T. & Koyama, M. Optuna: A Next-generation Hyperparameter Optimization Framework. Preprint at https://doi.org/10.1145/3292500.3330701 (2019).
3. G. Landrum. RDKit: Open-Source Cheminformatics Software. (2016).
4. Kühne, T. D. *et al.* CP2K: An Electronic Structure and Molecular Dynamics Software Package - Quickstep: Efficient and Accurate Electronic Structure Calculations. *J. Chem. Phys.* **152**, 194103 (2020).
5. Lu, T. A Comprehensive Electron Wavefunction Analysis Toolbox for Chemists, Multiwfn. *J. Chem. Phys.* **161**, 082503 (2024).
6. Momma, K. & Izumi, F. *VESTA 3* for Three-Dimensional Visualization of Crystal, Volumetric and Morphology Data. *J. Appl. Crystallogr.* **44**, 1272–1276 (2011).
7. Ong, S. P. *et al.* Python Materials Genomics (pymatgen): A Robust, Open-Source Python Library for Materials Analysis. *Comput. Mater. Sci.* **68**, 314–319 (2013).
8. Li, Q. *et al.* Harmonizing the Bilateral Bond Strength of the Interfacial Molecule in Perovskite Solar Cells. *Nat. Energy* **9**, 1506–1516 (2024).